\documentclass[a4paper]{article}

\usepackage{setspace}
\usepackage{fullpage}
\providecommand{\keywords}[1]{\textbf{\textit{Index terms---}} #1}
\usepackage{cite}
\usepackage[utf8]{inputenc}
\usepackage{graphicx}
\usepackage{float}
\usepackage{color, colortbl}
\usepackage{xcolor}
\usepackage{array}
\usepackage{multirow}
\usepackage{footnote}
\usepackage{cite}
\usepackage[normal]{threeparttable}
\usepackage{amsmath}
\usepackage{mathtools}
\usepackage{flushend}

\usepackage[linesnumbered,vlined,norelsize]{algorithm2e}

\usepackage{caption}
\usepackage{subcaption}

\usepackage{hyperref}
\makeatletter
\newcommand{\removelatexerror}{\let\@latex@error\@gobble}
\makeatother

\makeatletter
\def\blfootnote{\xdef\@thefnmark{}\@footnotetext}
\makeatother

\usepackage{lipsum}

\title{Traffic Differentiation in Dense Collision-free WLANs using CSMA/ECA}
  \author{Luis Sanabria-Russo and Boris Bellalta\\ Department of Information and Communication Technologies \\ Universitat Pompeu Fabra\\ Edifici Tanger
c/ Tanger, 122-140 08018 Barcelona, Spain \\ \texttt{<luis.sanabria,boris.bellalta>@upf.edu}}

\begin{document}
\maketitle

\begin{abstract}
The ability to perform traffic differentiation is a promising feature of the current Medium Access Control (MAC) in Wireless Local Area Networks (WLANs). The Enhanced Distributed Channel Access (EDCA) protocol for WLANs proposes up to four Access Categories (AC) that can be mapped to different traffic priorities. High priority ACs are allowed to transmit more often than low priority ACs, providing a way of prioritising delay sensitive traffic like voice calls or video streaming. Further, EDCA also considers the intricacies related to the management of multiple queues, virtual collisions and traffic differentiation. Nevertheless, EDCA falls short in efficiency when performing in dense WLAN scenarios. Its collision-prone contention mechanism degrades the overall throughput to the point of starving low priority ACs, and produce priority inversions at high number of contenders.

Carrier Sense Multiple Access with Enhanced Collision Avoidance (CSMA/ECA) is a compatible MAC protocol for WLANs which is also capable of providing traffic differentiation. Contrary to EDCA, CSMA/ECA uses a contention mechanism with a deterministic backoff technique which is capable of constructing collision-free schedules for many nodes with multiple active ACs, extending the network capacity without starving low priority ACs, as experienced in EDCA. This work analyses traffic differentiation with CSMA/ECA by describing the mechanisms used to construct collision-free schedules with multiple queues. Additionally, evaluates the performance under different traffic conditions and a growing number of contenders. Furthermore, it introduces a way to eliminate Virtual Collisions (VC), which also contribute to the throughput degradation in EDCA WLANs. Simulation tests are performed using voice and video packet sources that emulate commonly used codecs. Results show CSMA/ECA outperforming EDCA in different commonly-found scenarios with higher number of users, including when both MAC protocols coexist in the same WLAN.
\end{abstract}


%\doublespacing
\onehalfspacing
%\singlespace

\section{Introduction}
Wireless Local Area Networks (WLANs or WiFi networks~\cite{802Standards}) are a popular solution for wireless connectivity. Ranging from computers to wearable devices, it has widespread adoption. Unlike other wireless technologies, the medium in WLANs is shared. Every user having a packet to transmit must join a contention for the channel, whose winner will gain access and attempt a transmission. The Distributed Coordination Function (DCF) is based on Carrier Sense Multiple Access with Collision Avoidance (CSMA/CA)\footnote{DCF and CSMA/CA will be used interchangeably throughout the rest of the text.}, it coordinates access to the wireless channel in a completely distributed way by deferring each contender's transmission for a random backoff period.

WiFi's increasing adoption coupled with the envisioned multi-media, real-time, and bandwidth-hungry future use cases push the need for mechanisms to prioritise traffic in order to ensure Quality of Service (QoS) in dense scenarios with many nodes~\cite{HEW-scenarios,bellalta2015WCM}; i.e., to provide advantageous conditions for throughput or delay sensitive applications like video calls, streaming, or gaming. The Enhanced Distributed Channel Access (EDCA) (specified in IEEE 802.11e protocol~\cite{80211e}), builds over DCF in order to provide this kind of traffic differentiation.

EDCA proposes up to four queues or Access Categories (ACs), each one working as an instance of the DCF. Essentially, at the Medium Access Control (MAC) layer EDCA allows the higher priority ACs to access the channel more often. Traffic differentiation in EDCA is obtained by defining minimum and maximum contention windows (CW), and different waiting periods called Arbitration Inter-Frame Spacings (AIFS), which together essentially defer low priority traffic to make more transmission slots available for time-sensitive flows. Lastly, EDCA considers different Transmission Opportunities (TXOP) for each AC, which basically determines how much time an AC can allocate the channel upon each transmission attempt.

It is possible to adjust EDCA priority parameters\footnote{CW, AIFS and TXOP default values.} in the attempt to enhance the overall QoS in high priority ACs. The approach to this task varies. Algorithms can be completely distributed, like the EDCA recommendation itself, of centralised at the Access Point (AP). Centralised algorithms may use Call Admission Control (CAC)\footnote{also specified in IEEE 802.11e.} to accept or reject flows from stations and consequently announce MAC parameter adjustments. Nevertheless, as traffic differentiation is provided using a random backoff contention mechanism, collisions gravely impact EDCA, often resulting in priority inversions and throughput starvation of low priority ACs as the number of contenders in the WLAN increases. Distributed adaptation of contention parameters based on the current number of contenders, $N_{t}$ leverages this issue for EDCA~\cite{1208922,lopez-toledo2006aoi}, increasing the channel utilisation. Nevertheless, the type of source traffic negatively impacts the precision of the estimation, providing averages instead of precise number of contenders that could be used to provide optimised contention parameters.

Carrier Sense Multiple Access with Enhanced Collision Avoidance (CSMA/ECA)~\cite{sanabria2014high,research2standards} is capable of achieving greater throughput than CSMA/CA mostly due to its ability to make a more efficient use of the channel, particularly in dense WLAN networks~\cite{sanabria2014high}. Instead of deferring each contender's transmission for a random backoff period (as in CSMA/CA), CSMA/ECA instructs contenders to use a deterministic backoff after successful transmissions. In doing so, contenders that successfully transmitted in the past will do so again without colliding with other successful contenders in future cycles. By constructing collision-free schedules CSMA/ECA is able to increase the number of high priority flows supported, and to avoid the throughput starvation of low priority ACs (as experienced in EDCA~\cite{990806}), hence yielding greater throughput. Extensions to CSMA/ECA, namely Hysteresis and Fair Share, allow to dynamically increase the size of the deterministic backoff to create collision-free schedules for more transmitters, while still providing throughput fairness among users. Conversely, to avoid deterministic backoffs too big (which may increase the time between successful transmissions), CSMA/ECA incorporates the Schedule Reset (SR) mechanism. SR registers the state of each time slot\footnote{Empty or busy.} between backlogged ACs's successful transmissions, looking for opportunities to reduce the deterministic backoff. Analogous to EDCA, CSMA/ECA can be extrapolated to handle multiple queues, thus providing a way to implement EDCA-compatible collision-free traffic differentiation in WLANs.

Although several studies analyse the performance or extend CSMA/ECA to support many contenders~\cite{bellalta2009vtc,barcelo2011tcf,ECA-DEMO-INFOCOM14,research2standards,sanabria2014high}, this work is the first to incorporate traffic differentiation into the protocol with all of its extensions. We provide the following contributions:

\begin{itemize}
	\item Adaptation of CSMA/ECA to support multiple queues for collision-free traffic differentiation.
	\item Introduction of the Smart Backoff mechanism for eliminating Virtual Collisions (VC).
	\item First simulation results on throughput, collisions and delay for CSMA/ECA with four ACs.
	\item Implementation of a realistic non-saturation scenario, using well-known models for voice and video codecs.
	\item Evaluation of the coexistence and backwards compatibility with EDCA.
\end{itemize}

Results show that CSMA/ECA uses a more efficient collision avoidance mechanism, wastes less channel time recovering from failed transmissions, yields higher throughput, provides traffic differentiation, and creates the possibility of supporting more high priority flows for a higher number of contenders. Equally important, results show that CSMA/ECA is able to coexist with EDCA without gravely impacting this type of nodes's throughput.

An overview of the traffic differentiation techniques used with EDCA will be provided in Section~\ref{section2}. Then, we will present CSMA/ECA and its ability to provide traffic differentiation in Section~\ref{section3}. The performance evaluation is shown in Section~\ref{section4}, while we draw our conclusions in Section~\ref{section5}.

\section{Related Work}\label{section2}
Each node with a packet to transmit must join a contention for the channel. In CSMA/CA, nodes are deferred for a fixed period of idle-channel time and then for a random backoff period before attempting transmission. Because it only considers a single kind of traffic, the default contention parameters are the same for all contenders. 

In this section we present the traffic differentiation capabilities of EDCA as well as other compatible contention parameters adjustment techniques proposed by the research community.

\subsection{Enhanced Distributed Channel Access}\label{EDCA}
EDCA provides traffic differentiation by defining three parameters for each of the four ACs. First, by adjusting the Transmission Opportunity (TXOP) an AC may transmit several packets without repeating the contention for the channel, thus achieving greater throughput than other ACs. The other two parameters are related to the contention process, namely the Contention Window (CW$_{\min}$ and CW$_{\max}$, for minimum and maximum respectively) and the Arbitration Inter-Frame Spacing (AIFS). The contention windows limit the random backoff period, while the AIFS defines the fixed waiting period when the channel is idle. ACs with low contention windows and short AIFS will access the channel quicker, that is, have higher priority.

WLANs time is slotted. That is, it is composed of tiny empty slots of fixed duration $\sigma_{e}$, collisions and successful slots (which contain collisions or a successful transmission, their duration denoted by $\sigma_{c}$ and $\sigma_{s}$, respectively)\footnote{Empty slots are much shorter than collision or successful slots, that is, $\sigma_{e}\ll\min(\sigma_{c},\sigma_{s})$.}. DCF instructs backlogged stations to wait for a random number of empty slots (random backoff period) before attempting transmission. Transmissions always start at the beginning of a slot.

EDCA extends directly from DCF. In its place, EDCA declares up to four Access Categories (AC), each one an instance of DCF with different contention parameters that allow a statistical prioritisation among them~\cite{perahia2013next}. Traffic types, declared by the IEEE 802.1D standard~\cite{8021d} are then mapped to the four ACs in EDCA (MAC bridging). The mapping is shown in Table~\ref{tab:prioritiesMap}.

	\begin{table}[t]
		\centering
		\caption{AC relative priorities and mapping from 802.1D user priorities (extracted from~\cite{perahia2013next})}
		\label{tab:prioritiesMap}
		\begin{tabular}{|c|c|c|c|}
			\hline
			{\bfseries Priority} & {\bfseries 802.1D User priority} & {\bfseries AC} & {\bfseries Designation}\\
			\hline
			\multirow{2}{*}{Lowest} & 1 & \multirow{2}{*}{BK} & \multirow{2}{*}{Background}\\
			\cline{2-2}
							     & 2 &				    &\\
			\hline
			\multirow{2}{*}{}	     & 0 & \multirow{2}{*}{BE} & \multirow{2}{*}{Best Effort}\\
			\cline{2-2}
							     & 3 & 				     &\\
			\hline
			\multirow{2}{*}{}	     & 4 & \multirow{2}{*}{VI} & \multirow{2}{*}{Video}\\
			\cline{2-2}
							     & 5 & 				     &\\
			\hline
			\multirow{2}{*}{Highest}& 6 & \multirow{2}{*}{VO} & \multirow{2}{*}{Voice}\\
			\cline{2-2}
							     & 7 & 				     &\\
			\hline			
		\end{tabular}
	\end{table}
	
Every backlogged AC joins the contention for the channel by drawing a random backoff, $B[\text{AC}]\leftarrow\mathcal{U}[0,\text{CW}_{\min}[\text{AC}]-1]$; where CW$_{\min}[\text{AC}]$ is the minimum contention window for said AC. Each AC waits for a fixed $AIFS[\text{AC}]=SIFS+\sigma_{e}(AIFSN[\text{AC}]-1)$ period of inactivity in the channel and then starts decrementing its random backoff\footnote{The Short Inter-Frame Space (SIFS) is defined in~\cite{802Standards}. It is equal to 10 or 16$\mu$s for 802.11 n and ac/ax respectively.}. Each passing empty slot decrements $B[\text{AC}]$ in one. When the backoff counter expires ($B[\text{AC}] = 0$), the AC will attempt transmission. A successful transmission is declared upon the reception of an Acknowledgement (ACK) frame from the receiver, a collision is assumed otherwise. 

EDCA instructs the successful AC to reset its current contention window (CW$_{\text{curr}}[\text{AC}]$) to CW$_{\min}[\text{AC}]$, while failed transmissions force a retransmission attempt after doubling the current contention window, CW$_{\text{curr}}[\text{AC}]\leftarrow \min(2 * \text{CW}_{\text{curr}}[\text{AC}], \text{CW}_{\max}[\text{AC}])$. Table~\ref{tab:EDCAparams} shows the default CW, AIFSN and TXOP values specified for EDCA. 

	\begin{table}[t]
		\centering
		\caption{Deafult EDCA parameters (extracted from~\cite{perahia2013next})}
		\label{tab:EDCAparams}
		\begin{tabular}{|c|c|c|c|c|c|}
			\hline
			{\bfseries AC} & {\bfseries CW$_{\min}$} & {\bfseries CW$_{\max}$} &		{\bfseries m}		& {\bfseries AIFSN} & {\bfseries TXOP limit}\\
			\hline
			BK 		       & 		32			&		1024		   &			5			& 		8		  &		0 (only one MSDU)\\
			%\hline
			BE 		       & 		32			&		1024		   &			5			& 		4		  &		0 (only one MSDU)	\\
			%\hline
			VI 		       & 		16			&		32		  	   &			1			& 		3		  &		3.008 ms		\\
			%\hline
			VO 		       & 		8			&		16		  	   &			1			& 		3		  &		1.504 ms		\\
			%\hline
			Legacy	       & 		16			&		1024	  	   &			5			& 		3		  &		0 (only one MSDU)\\
			\hline
		\end{tabular}
	\end{table}

As it can be observed in Table~\ref{tab:EDCAparams}, ACs BK and BE may only send one MAC Service Data Unit (MSDU) upon each attempt. Whereas VI and VO can allocate the channel for longer periods. The TXOP parameter offers resource fairness rather than throughput fairness, that is, all ACs of the same category will receive close to the same average channel time regardless of its data rate. Furthermore, because the CW and AIFSN values for VI and VO are smaller than the others, on average these ACs will access the channel quicker; thus providing priority in the contention. 

%The relative AIFSN lengths for each AC are shown in Fig.~\ref{fig:AIFSN}.
%
%	\begin{figure}[t]
%	\centering
%		\includegraphics[width=0.7\linewidth]{figures/AIFSN.pdf}
%		\caption{AIFSN values for each AC. From lowest (BK) to highest priority (VO)}
%		\label{fig:AIFSN}
%	\end{figure}

While being effective in providing traffic differentiation and priority, in principle EDCA is unable to eliminate collisions. For instance, two ACs from different contenders may draw the same random backoff and consequently attempt transmission in the same time slot, causing a collision. Furthermore, if two or more ACs within a node experience a backoff expiration at the same instant, a Virtual Collision (VC) will occur. VC are resolved by granting the channel to the highest priority AC, while doubling the CW$_{\text{curr}}$[AC] for the lower priority ACs; just as it is done in the event of a real collision.

It follows directly from above that collisions waste channel time and thus contribute to the throughput degradation in WLANs. Moreover, the probability of collision increases as more contenders join the network, each one having four ACs attempting to gain access to the channel.

\subsection{EDCA enhanced}
Because ACs in EDCA perform contention independently of the others, each AC mimics a DCF station. This explains why the collision probability in EDCA is higher than in DCF networks with the same number of saturated nodes. Furthermore, the contention parameters in EDCA work better in scenarios with low number of contenders, but often cause starvation of low priority ACs in crowded scenarios (see~\cite{990806} and Section~\ref{sim:results}).

Great efforts have been directed towards parameter adjustments in EDCA, mostly to ensure QoS for high priority ACs while maintaining low delay and losses~\cite{throughputGuarantees,6614899,4594854}. For example, by dynamically adjusting the AIFS for each AC it is possible to maintain traffic differentiation while avoiding the starvation of low priority ACs. This is especially relevant in WLANs where all ACs are required to have effective throughput, like in~\cite{6614899}. Further, by randomising the AIFS values it is possible to increase the channel utilisation in EDCA~\cite{4594854}. 

MAC parameter adjustment algorithms work as functions that select future values for contention or transmission parameters in each AC. Most consider changing the contention windows, mainly because these were the only contention parameters in DCF. Nevertheless, adjustment of AIFS, and/or TXOP are also possible. These can be classified as~\cite{cano2010tuning}:
	\begin{itemize}
		\item Static or Adaptive: static algorithms define contention parameters for all ACs, which remain unchanged throughout the operation. An adaptive algorithm selects the best contention parameters for each AC depending on the detected flows. They also react to network congestion variations.
		\item Measurement or Model based: measurement-based algorithms divide time in periods, say $\Delta t$. By observing different metrics, e.g.: AC queue size, or collision rate during $\Delta t$ seconds, the algorithm estimates better MAC parameters to increase QoS in high priority ACs. On the other hand, model-based algorithms update MAC parameters every time a new flow is observed. These approaches can be combined, for instance, using Call Admission Control (CAC) coupled with a monitoring period of $\Delta t$. Such combination may accept or reject flows, and announce new MAC parameters according to the measured metrics, like~\cite{bellalta290call}.
		\item Centralised or Distributed: a key characteristic of EDCA, and DCF before it, is its distributed nature. That is, EDCA defines a static, measurement-based algorithm that reacts to network conditions. MAC parameter computation in distributed algorithms is performed at each node, independently. Centralised algorithms, additionally, make use of information obtained by a centralised entity, like the AP. Distributed algorithms do not need additional control messages to adjust MAC parameters, as opposed to centralised ones.
	\end{itemize} 

Another example of adaptive, distributed and measurement-based algorithms for WLANs is proposed in~\cite{1200574}. It follows EDCA rules for updating the CW after failed transmissions. Nevertheless, after a successful transmission the CW$_{\text{curr}}[\text{AC}]$ is \emph{slowly} reset to CW$_{\min}[\text{AC}]$ by computing a Multiplication Factor (MF), which itself depends on the ratio between failed transmissions and transmission attempts. This \emph{Slow Reset} of the CW$_{\text{curr}}[AC]$ reduces the collision probability of the immediate attempt after a successful transmission. In~\cite{Ksentini}, distributed TXOP adaptation is combined with a CAC. Called Enhanced TXOP (ETXOP), this algorithm estimates the network congestion using the number of times a station's backoff counter is frozen, and then adjusts TXOP sizes so the application's requirements for each AC are met. Combined with a distributed model-based algorithm, namely a CAC which handles flows coming from applications at each node, EXTOP ensures that only flows with a guaranteed QoS are accepted for contention.

Centralised algorithms may take advantage of an AP's point of view of the network and of its ability to transmit MAC parameter updates in beacon frames. In~\cite{bellalta290call}, a centralised CAC algorithm distinguishes between VoIP and TCP flows, and at the same time between uplink and downlink traffic. Measuring each flow type, required bandwidth, and average frame length the CAC reacts to each new flow request, adjusting CW, AIFS or TXOP values to comply with defined VoIP requirements, like delay and frame loss. The CAC handles the flows differently depending on its characteristics:
	\begin{itemize}
		\item Downlink TCP flow: if the number of existing downlink TCP flows is below an estimated threshold, the flow is admitted.
		\item Uplink TCP flow: if the number of existing uplink TCP flows is below an estimated threshold, the flow is admitted. Otherwise, other CW$_{\min}$ values are proposed via a Beacon frame so the newly arrived flow can be admitted.
		\item Downlink VoIP flow: if the number of packets in the queue for other downlink VoIP flows is below a threshold, the new flow is accepted and the threshold updated.
		\item Uplink VoIP flow: the flow is admitted if the grade of service of existing flows is not affected. On the positive case, other CW$_{\min}$ and AIFS values are proposed to admit the newly arrived flow.
		\item If no other parameter update is feasible, the flow is rejected.
	\end{itemize}

Algorithms may be combined, or focus on iteration in order to provide advantageous conditions for high priority traffic. Nevertheless, as proposals deviate too much from the IEEE 802.11 MAC standard, the chances of being accepted as an amendment decreases~\cite{WMP,perahia2008ieee}. Moreover, performance evaluations should implement updated audio and video source models, using specifications of widely-used codecs in order to mimic realistic scenarios~\cite{van2008traffic,menth2009source}.

The way traffic differentiation is defined in IEEE 802.11e is through a static, completely distributed, and measurement-based algorithm, that is, EDCA. As its goal is to provide QoS to high priority ACs, low priority traffic is often deferred to the point of throughput starvation. Additionally, EDCA's random backoff mechanism is prone to an elevated number of real and virtual collisions, widening the effects of throughput starvation to higher priority ACs\footnote{Throughput starvation is first observed in AC[BK], and then in AC[BE] as the number of contenders increases.}. 

\subsection{Collision-free WLANs}
Time slot reservation techniques are known to provide higher throughput and QoS in TDMA schemes, like LTE~\cite{canoLTEcoexistence}. By applying similar concepts (like organising transmissions according to a predefined schedule) to a completely decentralised CSMA, it is possible to reach collision-free operation. Using a Semi-Random Backoff (SRB)~\cite{HE} after successful transmissions, it is possible to achieve collision-free operation for high number of contenders. Other proposals, like ZC-MAC~\cite{ZMAC} and L-MAC\cite{L_MAC} define virtual cycles known to all users, in which stations allocate transmission slots. The selection of the same slot in future cycles is conditioned to the observed failed transmissions during the past cycle. These are examples of decentralised MAC protocols for WLANs that use the concept of slot reservation to provide collision-free operation.

\subsubsection{Zero Collision MAC}
Zero Collision MAC (ZC-MAC)~\cite{ZMAC} allows contenders to reserve one empty slot from a predefined virtual schedule of $M$-slots in length. If two or more stations select the same transmission slot, the involved contenders select randomly and uniformly other empty slot from those detected empty in the previous cycle plus the slot where they collided. Collision-free operation is achieved when all $N$ stations select different transmission slots in the schedule.

ZC-MAC is able to outperform CSMA/CA under different scenarios. Nevertheless, given that the length of ZC MAC's virtual cycle has to be predefined without actual knowledge of the real number of contenders in the deployment, the protocol is unable to provide a collision-free schedule when $N>M$. Furthermore, if $M$ is overestimated ($M\gg N$), the fixed-width empty slots between each contender's successful transmission are no longer negligible and contribute to the degradation of the network performance. Additionally, ZC-MAC nodes require common knowledge of where the virtual schedule starts/ends. This is not considered in CSMA/CA. Further, multiple queues or traffic differentiation are not considered.

\subsubsection{Learning-MAC}
Learning-MAC (L-MAC) and a survey of other collision-free MAC protocols for WiFi are presented in~\cite{L_MAC}. As its name suggests, L-MAC uses learning techniques to achieve collision-free schedules. It defines a \emph{learning strength} parameter, $\beta\in(0,1)$. Each contender starts by picking a slot $s$ for transmission from schedule $n$ of length $C$ at random with uniform probability. After transmission on slot $s(n)$, the node conditions the selection of the same slot in the next cycle according to the result of the transmission. (\ref{success}) and (\ref{collisions-eq}) extracted from~\cite{L_MAC} show the probability of selecting the same slot $s(n)$ in cycle $n+1$.

\begin{equation} \label{success}
		\left. \begin{aligned}
			p_{s(n)}(n+1)&=1,\\
			p_{j}(n+1)&=0,
		\end{aligned}
	\right\}
	\qquad \text{\emph{Success}}
\end{equation}
\begin{equation} \label{collisions-eq}
	\left. \begin{aligned}
			p_{s(n)}(n+1)&=\beta p_{s(n)}(n),\\
			p_{j}(n+1)&=\beta p_{j}(n)+\frac{1-\beta}{C -1},
		\end{aligned}
	\right\}
	\qquad \text{\emph{Collision}}
\end{equation}
\\
for all $j\neq s(n),~j\in \{1,\dots ,C\}$. That is, if a station successfully transmitted in $s(n)$, it will pick the same slot on the next schedule with probability one. Otherwise, it follows~(\ref{collisions-eq}).

The selection of $\beta$ implies a compromise between fairness and convergence speed, which the authors determined $\beta=0.95$ to provide satisfactory results.

L-MAC converges to collision-free schedules in a few cycles. Further extensions to L-MAC introduced an \emph{Adaptative} schedule length in order to increase the number of supported contenders in a collision-free schedule. This adaptive schedule length is doubled or halved depending on the presence of collisions or many empty slots per schedule, respectively. As ZC-MAC, L-MAC stations require common knowledge of the start/end of the schedule. 

\subsubsection{Descentralised collision-free traffic differentiation}
These reservation-like protocols, namely, L-MAC and ZC-MAC could be adapted to support traffic differentiation by using multiple schedules. Semi-Random Backoff~\cite{HE} is able to build collision-free schedules using a deterministic backoff after successful transmissions. Further, SRB proposes traffic differentiation using different deterministic backoffs for each AC. Nevertheless, frame aggregation techniques are not considered, leading to throughput unfairness issues. Moreover, results for non-saturated scenarios do not follow realistic traffic sources for voice and video, providing inaccurate modelling of nodes' behaviour regarding the arrival or withdrawal from contention. Finally, as backwards compatibility is considered a key aspect of WiFi's popularity, evaluations should include evaluations of mixed scenarios using accurate models for traffic sources.

%Nevertheless, all these approaches either require information about the number of nodes participating in the contention, use a centralised entity (often the AP), or inject additional traffic to the network, which may be unsuitable for crowded scenarios.
\section{Traffic Differentiation with CSMA/ECA}\label{section3}
Carrier Sense Multiple Access with Enhanced Collision Avoidance~\cite{sanabria2014high, research2standards} is able to build collision-free schedules by using a deterministic backoff after successful transmissions. That is, if all saturated contenders are able to perform a successful transmission and then pick a deterministic backoff, they will not collide among each other in future transmissions.
	
When a packet arrives at an empty MAC queue, stations generate a random backoff $B\leftarrow\mathcal{U}[0,\text{CW}_{\min}-1]$, just as in DCF. Every passing empty slot decrements $B$ in one. When $B=0$, the station will attempt a transmission. If the transmission fails, the node will increment its backoff stage $k\in[0,m]$ in one (where $m$ is the maximum backoff stage of typical value $m=5$) and use another random backoff $B\leftarrow\mathcal{U}[0,\text{CW}(k)-1]$; where $\text{CW}(k)=2^{k}\text{CW}_{\min}$ is the contention window at backoff stage $k$. Otherwise, the successful station will then pick a deterministic backoff for its next transmission, $B_{\text{d}}\leftarrow \lceil \text{CW}_{\min}/2\rceil-1$. This value of $B_{\text{d}}$ is roughly equal to the expectation of a random backoff at the same backoff stage ($k=0$, in this case), making it fair and compatible with CSMA/CA stations~\cite{research2standards}.

\subsection{Supporting many more contenders with Hysteresis, Fair Share and the Schedule Reset mechanism}\label{scheduleReset}

CSMA/ECA is also capable of allocating many contenders in a collision-free schedule by not reseting the backoff stage $k$ after a successful transmission, as opposed to CSMA/CA. That is, a node at backoff stage $k$ would select $B_{\text{d}}\leftarrow \lceil \text{CW}(k)/2\rceil-1$ as its deterministic backoff after a successful transmission. This extension to CSMA/ECA is called Hysteresis. 

Hysteresis forces some contenders to have larger schedules than others, resulting in an unfair distribution of the network resources. This effect can be compensated by allowing nodes at backoff stage $k$ to transmit $2^{k}$ frames, performing MPDU aggregation (AMPDU) and using Block Acknowledgement~\cite{perahia2013next} upon each transmission attempt. We call this extension Fair Share and it ensures an even distribution of the available throughput among contenders. CSMA/ECA is able to outperform CSMA/CA, mainly due to the more efficient collision avoidance mechanism and the aggregation technique suggested by Fair Share.

CSMA/ECA instructs nodes not to reset their backoff stage after a successful transmission. This is done in order to increase the cycle length and provide a collision-free schedule for many contenders, which is desirable in dense scenarios. Nevertheless, having a big deterministic backoff increases the time between successful transmissions. Furthermore, if not operating in a scenario with many nodes the empty slots between transmissions are not longer negligible and degrade the overall throughput of the system. For instance, if nodes withdraw from the contention their previously used slots will now be empty. Contenders should be aware of this issue and pursue opportunities to reduce their deterministic backoff without sacrificing too much in collisions, or having any precise knowledge about the number of contenders.

The \emph{Schedule Reset} mechanism introduced in~\cite{sanabria2014high} consists on finding the smallest CSMA/ECA collision-free schedule between a contender's transmissions and then change the node's deterministic backoff to fit in that schedule. Take a contender with a $B_{\text{d}}=31$ as an example. By listening to the slots between its transmissions, the node should be able to determine the availability of smaller (and possibly) collision-free schedules. 

\begin{figure}[t!]
	\centering
		\includegraphics[width=0.8\linewidth]{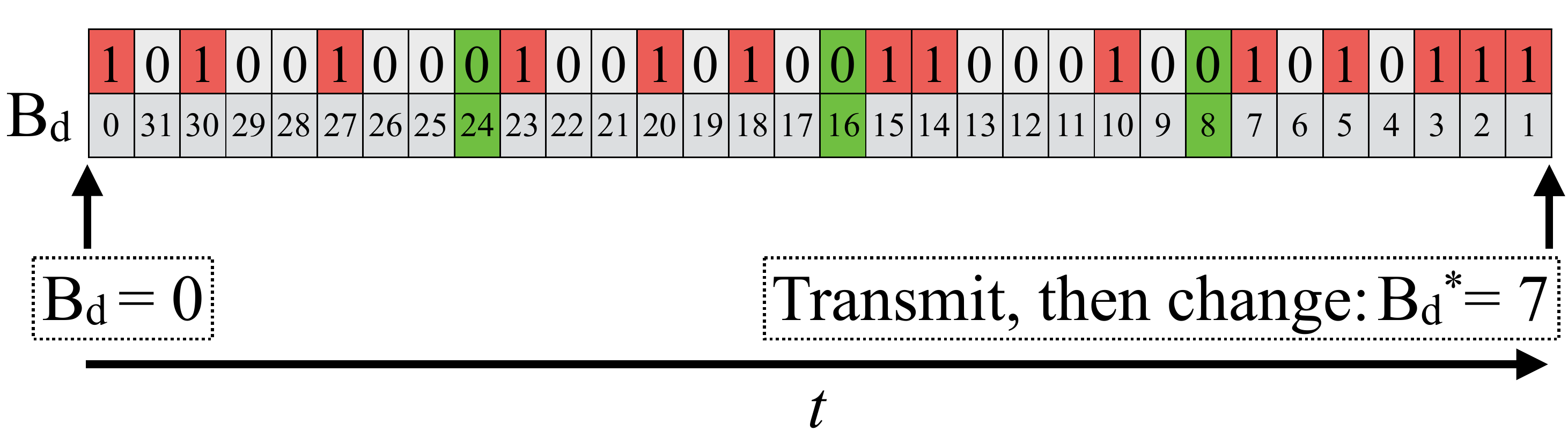}
		\caption{Example of the Schedule Reset mechanism (extracted from~\cite{sanabria2014high})}
		\label{fig:scheduleReset1}
	\end{figure}

Figure~\ref{fig:scheduleReset1} shows the slots between the transmissions of a contender with $B_{\text{d}}=31$. Starting from the left, the current $B_{\text{d}}=0$ means that the slot will be filled with the contender's own transmission. Each following slot containing either a transmission or a collision is identified with the number one, while empty slots are marked with a zero. Notice that the highlighted empty slots appear every eight slots, suggesting that a schedule reduction from $B_{\text{d}}=31$ to $B_{\text{d}}^{*}=7$ is possible\footnote{With $CW_{\min}=16$, the change of $B_{d}= \lceil CW(k)/2\rceil-1$ simply represents a reduction of the backoff stage $k$. Specifically, from $k=2$ to $k=0$.}. The schedule change is performed after the contender's next successful transmission. 

Schedule Reset (SR) is implemented in CSMA/ECA by filling a bitmap $b$ of size $B_{\text{d}}+1$. Each bit $t,~t\in\{0,\ldots ,B_{\text{d}}\}$ in the bitmap is the result of a bitwise OR operation between its current value, $b[t]$ and the state of the observed slot; which equals to one when busy or zero when idle. After $\gamma$ consecutive successful transmissions (sxTx), the bitmap is evaluated. If a change of schedule is possible, it is executed just after the next successful transmission.

It is possible to configure Schedule Reset in two modes, namely \emph{conservative} and \emph{aggressive}. These modes relate to the number of consecutive transmissions needed to evaluate the bitmap, that is, $\gamma$. A conservative SR contains the information of all users' transmissions, therefore no additional collisions are introduced as a consequence of the schedule change\footnote{Assuming perfect channel conditions and saturated sources.}. This implies a value of $\gamma=2^{(m-k)+1}$. On the other hand, setting $\gamma=1$ triggers a bitmap evaluation after just two consecutive transmissions, rendering this choice of $\gamma$ the aggressive mode.

\begin{figure}[tb]
	\centering
		\includegraphics[width=0.35\linewidth, angle=-90]{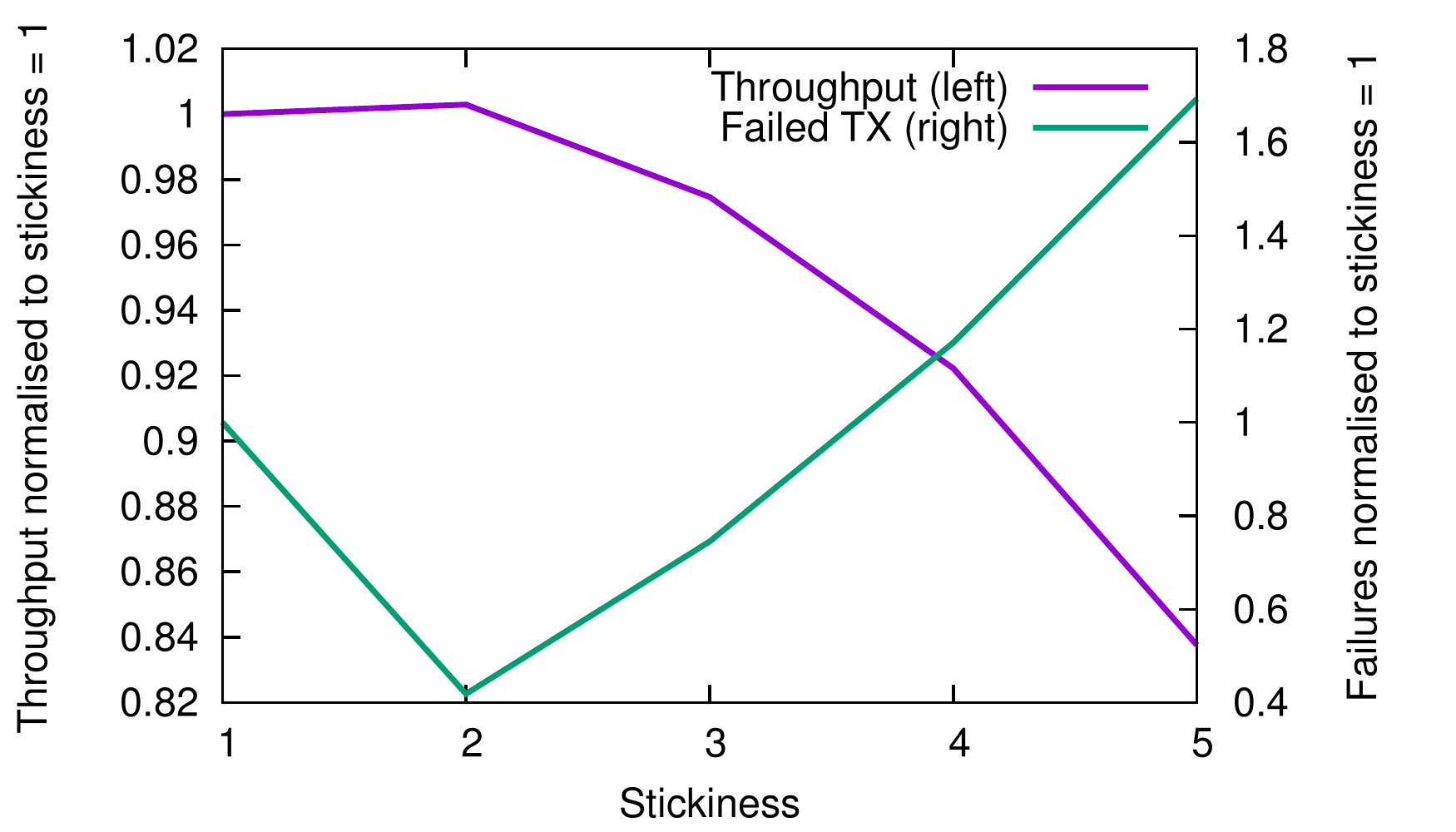}
		\caption{CSMA/ECA average throughput and failed transmission with different levels of stickiness, fixing the number of contenders $N=20$ in saturated traffic conditions (as explained in Section~\ref{subsect:simParams})}
		\label{fig:stickEv-throughput-overallOnly}
	\end{figure}

Aggressive Schedule Reset coupled with an increase in the Stickiness after an effective schedule change has proven to be suitable for noisy scenarios in real hardware implementations of CSMA/ECA~\cite{sanabria2014high}. This work uses the same settings to provide the simulation results in Section~\ref{section4}. Stickiness is not a new concept to CSMA/ECA~\cite{barcelo2011tcf}. It simply instructs the contender to \emph{stick} to the deterministic backoff even in the event of \emph{stickiness} number of failed transmissions. This allows for a faster convergence towards a collision-free schedule. CSMA/ECA with a default level of stickiness equal to $1$ has proven to provide the better combination of high throughput and low collisions, as shown in Figure~\ref{fig:stickEv-throughput-overallOnly}. 

To summarise, simulations results presented in Section~\ref{section4} use Aggressive Schedule Reset and increase from ${\texttt{stickiness}}=1$ to ${\texttt{stickiness}}=2$ after a successful reduction of the schedule. This is called Schedule Reset's Dynamic Stickiness. 

\subsection{Incorporating multiple ACs into CSMA/ECA: CSMA/ECA$_{\text{QoS}}$}
CSMA/ECA and its extensions are able to construct collision-free schedules under saturated conditions, outperforming CSMA/CA. Furthermore, CSMA/ECA uses the same default contention parameters as CSMA/CA, so the compatibility is maintained~\cite{sanabria2014high}.

Providing priority is to ensure a more frequent access to some ACs over others. In CSMA/ECA this is only subject to the deterministic backoff. That is, an AC using a shorter $B_{\text{d}}$ would in turn access the channel more often than those using a larger one. To maintain compatibility with EDCA, CSMA/ECA considers the same four ACs.

\begin{figure}[tb]
	\centering
		\includegraphics[width=0.7\linewidth]{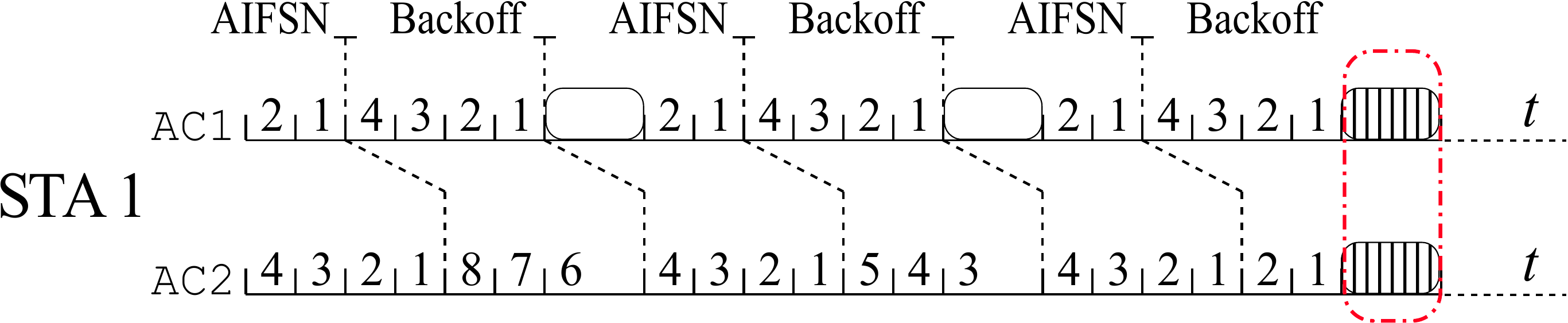}
		\caption{Example temporal evolution of CSMA/ECA with two ACs using AIFS resulting in a virtual collision. Only considering AIFSN values of 2 and 4, $B_{\text{d}}$ of 4 and 8 for AC1 and AC2 respectively}
		\label{fig:AIFSinECA}
	\end{figure}

Nevertheless, AIFS and TXOP are not fit for multiple CSMA/ECA queues. For instance, AIFS values are not required since differentiation is only provided by the deterministic backoff. The incorporation of different AIFS for each category would trigger Virtual Collisions that in turn may disrupt an existent collision-free schedule with real collisions. Figure~\ref{fig:AIFSinECA} shows a VC in CSMA/ECA with two queues (indicated by the outline) consequence of using AIFS during a collision-free schedule. As the lower priority AC proceeds to select a random backoff, its next transmission may disrupt any ongoing collision-free operation.

TXOP in EDCA ensures that all traffic from the same category receives on average the same channel time. In contrast, CSMA/ECA's goal through Fair Share is to provide close to equal average throughput to same-priority ACs. The combination of Fair Share and Schedule Reset provides throughput fairness through aggregation. Further, it attempts to evenly distribute the channel time among AC increasing the frequency of transmissions, permanently seeking opportunities to reduce the schedule. In order to provide a fair comparison with EDCA, Section~\ref{section4} also shows simulation results for CSMA/ECA using the default TXOP.

As EDCA extends DCF into four ACs, similarly there is an instance of CSMA/ECA for each AC. We will refer to CSMA/ECA with multiple ACs as CSMA/ECA$_{\text{QoS}}$ from here forward. Table~\ref{tab:newQoSparams} shows the CW, lowest and largest $B_{\text{d}}$, and maximum backoff stage $m$.

	\begin{table}[t]
		\centering
		\caption{CSMA/ECA$_{\text{QoS}}$ contention parameters for the simulations}
		\label{tab:newQoSparams}
		\begin{tabular}{|c|c|c|c|c|c|c|}
			\hline
			{\bfseries AC} & {\bfseries CW$_{\min}$} & {\bfseries CW$_{\max}$} & {\bfseries m} & {\bfseries lowest $B_{\text{d}}$} & {\bfseries highest $B_{\text{d}}$}\\
			\hline
			BK		       &	32				&		1024		  & 		5	&			15		        &		511		\\
			BE		       &	32				&		1024		  &		5	&			15		        &		511		\\
			VI		       &	16				&		512		  & 		5	&			7		     	&		255		\\
			VO		       &	8				&		256		  & 		5	&			3		        &		127		\\
			Legacy	       &	32				&		1024		  & 		5	&			15		        &		511		\\
			\hline
		\end{tabular}
	\end{table}

\begin{figure*}[tb]
	\centering
		\includegraphics[width=\linewidth]{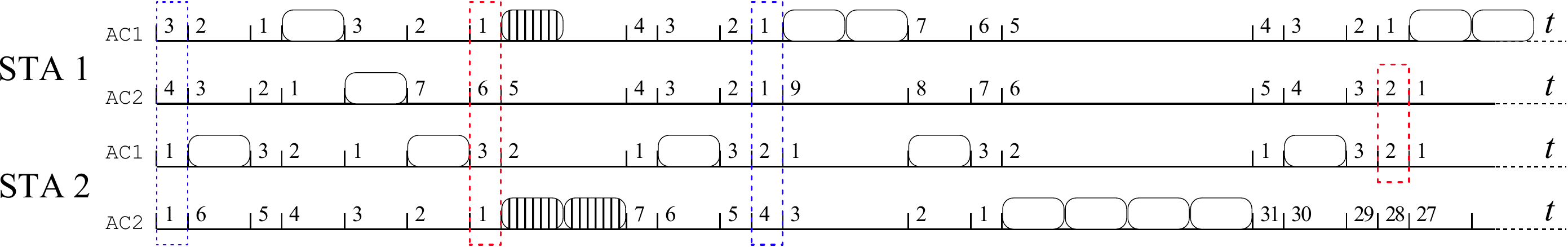}
		\caption{An example of the temporal evolution of CSMA/ECA$_{\text{QoS}}$ in saturation ($CW_{\min}[\text{AC1,AC2}]=[8,16]$; $m[\text{AC1,AC2}]=[5,5]$)}
		\label{fig:ecaQoS}
	\end{figure*}

Figure~\ref{fig:ecaQoS} shows an example of CSMA/ECA$_{\text{QoS}}$ with two contenders and two ACs; where AC1 has higher priority than AC2. In the figure, the first outline indicates a VC between AC1 and AC2 from STA-2. VC in CSMA/ECA$_{\text{QoS}}$ are handled just as in EDCA, that is, the AC with the highest priority is granted access to the channel, while the other ACs involved in the VC double their contention windows and use a random backoff for the next transmission. Consequently, AC1 from STA-2 successfully transmits and then uses $B_{\text{d}}=\frac{2^{0}CW_{\min}[\text{AC1}]}{2}-1= 3$.

Still on Figure~\ref{fig:ecaQoS}, the second outline indicates a collision between STA-2's AC2 and AC1 from STA-1. At this moment in time STA-2 AC2's backoff stage has been increased in two occasions ($k[\text{AC2}]=2$). When said AC2 is able to transmit, it sends $2^{k[\text{AC2}]}$ packets according to Fair Share. Then, it uses a deterministic backoff, $B_{\text{d}}=\frac{2^{k[\text{AC2}]}CW_{\min}[\text{AC2}]}{2}-1=31$. The third outline in Figure~\ref{fig:ecaQoS} indicates an VC in STA-1, which is resolved allowing AC1 and deferring AC2's transmission using a random backoff with a doubled CW. A future collision between STA-2's AC1 and AC2 from STA-1 is highlighted by the last outline.

\subsection{Collisions and Virtual Collisions-free operation using Smart Backoff}\label{ECAqosCollisionFree}
Consider a complete schedule of length $C=2^{m}\text{CW}_{\min}$, and $m=5$. With CSMA/ECA and a single AC is possible to allocate a collision-free transmission slot for up to $C/2=512$ contenders (the highest $B_{\text{d}}+1$ for AC Legacy in Table~\ref{tab:newQoSparams}). Nevertheless, with CSMA/ECA$_{\text{QoS}}$ and all ACs in saturation i.e., have a packet to transmit, each contender mimics the behaviour of four saturated CSMA/ECA nodes (one for each AC). This means that the total number of supported collision-free contenders will be reduced in order to provide a transmission slot for all the ACs in the network. If all the ACs are in saturation, CSMA/ECA$_{\text{QoS}}$ can provide collision-free operation for up to $2^{(m[\text{VO}]-3)}\text{CW}_{\min}[VO] = 32$ contenders, where $m[\text{VO}]$ is the maximum backoff stage of the AC with the smallest CW$_{\max}$, that is AC[VO] in Table~\ref{tab:newQoSparams}\footnote{The maximum number of collision-free contenders in saturation is reduced when using the Schedule Reset Mechanism. This is due to the reduction of the average backoff stage of AC[VO], $k[VO]\leq m[VO]$.}. 

%Even-though Table~\ref{tab:ecaQosParams} provides CSMA/ECA$_{\text{QoS}}$'s default contention parameters, these will be adjusted in Section~\ref{sim:results} in order to support more contenders while in saturation.

VCs in CSMA/ECA$_{\text{QoS}}$ force lower priority ACs to defer their transmissions using a random backoff. Therefore, VCs can disrupt any existent collision-free schedule in CSMA/ECA$_{\text{QoS}}$, wasting channel time recovering from collisions and degrading the overall throughput. Given that all AC's backoff counters are known to the contender, there is nothing preventing it from using this information to avoid future VCs.

CSMA/ECA$_{\text{QoS}}$ eliminates VCs by picking a $B[\text{AC}]$ that is not equal to any of the other AC's counters. This is achieved by selecting a number whose absolute difference with each of the other AC's counters is not a multiple of each comparison's smallest deterministic backoff. Algorithm~\ref{alg:smartBackoff} decribes the process of selecting what is referred to as a \emph{Smart Backoff} in CSMA/ECA$_{\text{QoS}}$. It shows four ACs, although it can used to eliminate VCs with as many ACs as needed. Smart Backoff is used instead of a random backoff in CSMA/ECA$_{\text{QoS}}$, regardless of the aggregation mechanism used.

%This is exemplified in Figure~\ref{fig:smartBackoff1}. The figure shows a contender with four ACs and $B_{\text{d}}[4]=[4,8,16,16]$, for AC1 to AC4 respectively. AC1 effectively transmits, but AC3 suffers from a VC. It then selects a random backoff, $B[\text{AC3}]=24$, which is different from any other AC's counters but do not consider the deterministic backoff selected by successful ACs. To eliminate future VCs in CSMA/ECA$_{\text{QoS}}$ the successes of other ACs's transmissions should also be taken into account. 

\begin{center}
\begin{minipage}{0.7\linewidth}
	\begin{algorithm}[H]
		$AC:=4$\tcp*[l]{number of Access Categories}
		$CW_{\min}[AC]$\tcp*[l]{$CW_{\min}$ for all ACs}
		$B_{\text{d}}[AC]$\tcp*[l]{$B_{\text{d}}$ for all ACs}
		$B[AC]$\tcp*[l]{current $B$ from all ACs}
		$k[AC]$\tcp*[l]{current backoff stage}
		$F[AC]:=\{0\}$\;
		$Cb[AC]:=\{0\}$\;
		\tcp{}\tcp{looking for a suitable $B[i]$; $i\in[1,AC]$}\tcp{}
		\While{$($F$~\neq 1)$ or $($Cb$~\neq 1)$}	
		{
			$B[i]\gets\mathcal{U}[0,2^{k[i]}{CW_{\min}[i]}]$\;
			\For{$(j = 1; j\leq \text{AC};j++)$}
			{
				\If{$(j\neq i)$}{
					$F[j]\gets |B[i]-B[j]|$~mod~$[\min(B_{\text{d}}[i],B_{\text{d}}[j])]$\;
					\If{$(F[j]\neq 0)$} 
					{	
						$F[j]\gets 1$\;
					}
					\eIf{$(B[i]\neq B[j])$}{
						$Cb[j]\gets 1$\;
					}{
						$Cb[j]\gets 0$\;
					}
				}
			}
		}
		\Return $(B[i])$\;
		\vspace{0.2cm}
		\caption{Smart Backoff: eliminating Virtual Collisions in CSMA/ECA$_{\text{QoS}}$}
		\label{alg:smartBackoff}
	\end{algorithm}
\end{minipage}
\end{center}

What results from Algorithm~\ref{alg:smartBackoff} is a Smart Backoff counter that will not cause a VC on the next transmission attempts.

%	\begin{table}[tb]
%		\centering
%		\caption{Deafult CSMA/ECA$_{\text{QoS}}$ contention parameters}
%		\label{tab:ecaQosParams}
%		\begin{tabular}{|c|c|c|c|c|c|}
%			\hline
%			{\bfseries AC} & {\bfseries CW$_{\min}$} & {\bfseries CW$_{\max}$} & {\bfseries m} & {\bfseries lowest $B_{\text{d}}$} & {\bfseries highest $B_{\text{d}}$}\\
%			\hline
%			BK		       &	32				&		1024		  & 		5	&			15		        &		511\\
%			BE		       &	32				&		1024		  &		5	&			15		        &		511\\
%			VI		       &	16				&		32			  & 		1	&			7		        &		15\\
%			VO		       &	8				&		16		 	 & 		1	&			3		        &		7\\
%			Legacy	       &	32				&		1024		  & 		5	&			15		        &		511\\
%			\hline
%		\end{tabular}
%	\end{table}

%	\begin{figure}[tb]
%	\centering
%		\includegraphics[width=0.5\linewidth]{figures/smartBackoff1.pdf}
%		\caption{Four AC backoff counters in a CSMA/ECA$_{\text{QoS}}$ node with $B_{\text{d}}[4]=[4,8,16,16]$. After a VC or collision, the new random backoff counter must not be equal to any other AC's counter}
%		\label{fig:smartBackoff1}
%	\end{figure}
\section{Performance Evaluation}\label{section4}
In order to test the traffic differentiation in CSMA/ECA$_{\text{QoS}}$ and its capability of outperform EDCA in terms of number of supported delay-sensitive flows and aggregate throughput, we have used a customised version of the COST simulator~\cite{COST}, which is available via~\cite{CSMA-ECA-HEW}. If not expressed otherwise, each point in the presented figures is obtained from averaging twenty executions of duration equal to forty seconds. Further considerations:
	\begin{itemize}
		\item PHY/MAC headers, and other unspecified parameters follow the IEEE 802.11ax ($5$~GHz) standard~\cite{IEEE80211ax}.
		\item All nodes can be assumed to be in communication range with each other.
		\item Transmission of several frames per attempt supposes AMPDU aggregation with compressed Block ACK~\cite{perahia2013next}.
		\item RTS/CTS mechanism is used, as transmitting multiple frames in a TXOP requires a protection mechanism in EDCA~\cite{80211e}.
%		\item CSMA/ECA$_\text{QoS}$ uses higher CW$_{\max}$ values in order to achieve collision-free schedule with higher number of contenders (see Table~\ref{tab:newQoSparams}).
		\item Smart Backoff is used in CSMA/ECA$_\text{QoS}$.
		\item Aggressive Schedule Reset is used, with $\gamma=1$.
		\item Dynamic Stickiness defines a maximum ${\texttt{stickiness = 2}}$.
		\item CSMA/ECA$_{\text{QoS}}$ AC[BK] does not use Schedule Reset in order to provide differentiation with AC[BE].
	\end{itemize}

The RTS/CTS message exchange between transmitter/received was originally intended to solve the hidden node problem in WLANs~\cite{perahia2013next}. However, it also has advantages for a large number of contenders, as it reduces the collision duration, which compensates for the RTS/CTS overhead. Initially, a transmitter enters in contention in order to send a short Request to Send (RTS) message to the receiver. Consequently, the receiver performs contention to respond with a Clear to Send (CTS) message (which is received by all contenders), allocating the next TXOP to the transmitter. Upon reception of the CTS message, the transmitter is granted contention free access to the channel during TXOP. Collisions using the RTS/CTS mechanism are shorter that using Basic Access (BA) (in which collisions are normally assumed to occupy as much channel time as a successful transmission), given the short size of RTS and CTS packets.

Additionally, Table~\ref{tab:mac-params} provides information about relevant PHY and MAC parameters used in the simulator.
	\begin{table}[t]
		\centering
		\caption{PHY and MAC parameters for the simulations~\cite{CSMA-ECA-HEW}}
		\label{tab:mac-params}
		\begin{tabular}{|c|c|}
			\hline
			\multicolumn{2}{|c|}{{\bfseries PHY}}\\
			\hline
			{\bfseries Parameter} & {\bfseries Value}\\
			\hline
			PHY rate & 65~Mbps\\
			Channel Width & 20~MHz\\
			Number of Streams & 1\\
			OFDM bits/symbol & 6\\
			Coding rate & 3/4\\
			Empty slot & $9~\mu s$\\
			DIFS & $34~\mu s$\\
			SIFS & $16~\mu s$\\
			\hline
			\multicolumn{2}{|c|}{{\bfseries MAC}}\\
			\hline
			{\bfseries Parameter} & {\bfseries Value}\\
			\hline
			Maximum retransmission attempts & 7\\
%			BE and BK default Packet size (Bytes) & 1470\\
			MAC queue size (Packets) & 1000\\
			\hline
			\multicolumn{2}{|c|}{{\bfseries CSMA/ECA$_\text{QoS}$}}\\
			\hline
			{\bfseries Parameter} & {\bfseries Value}\\
			\hline
			Schedule Reset mode & aggressive ($\gamma=1$)\\
			Dynamic stickiness & on\\
			Smart Backoff & on\\
			\hline
		\end{tabular}
	\end{table}
	
	\begin{table}[t]
		\centering
		\caption{Traffic sources detail: 1) AC[VO]: Internet Low Bit Rate Codec (iLBC)~\cite{andersen2004internet} source settings for the non-saturation scenario. Following a geometric distribution of talkspurts and silence intervals. Durations follow \emph{Geom-APD-W0} settings in~\cite{menth2009source}. 2) AC[VI]: H.264/AVC source settings for the non-saturation scenario. Using GOP composed of 3 B frames per I/P frames (G16-B3)~\cite{menth2009source}. 3) AC[BE] and AC[BK].}
		\label{tab:voice}
		\begin{tabular}{|c|c|}
			\hline
			\multicolumn{2}{|c|}{{\bfseries 1) AC[VO]}}\\
			\hline
			{\bfseries Parameter} & {\bfseries Value}\\
			\hline
			%Voice Activity Factor ($\alpha$) & $0.488$\\
			On duration & $3.110$~s\\
			Off duration & $3.2727$~s\\
			Rate & $15.2$~kbps\\
			Payload & $38$~B\\
			\hline
			\multicolumn{2}{|c|}{{\bfseries 2) AC[VI]}}\\
			\hline
			PSNR & $43.5$~dB, best\\
			GOP size & 16\\
			GOP & \emph{IBBBPBBBPBBBPBBB}\\
			Average I size & $5658$~B\\
			Average P size & $1634$~B\\
			Average B size & $348$~B\\
			Frame size standard deviation & 2 times the average\\
			Average Rate & $300$~kbps\\
			\hline
			\multicolumn{2}{|c|}{{\bfseries 3) AC[BE] and AC[BK]}}\\
			\hline
			Rate & 65 Mbps\\
			Payload & 1470 B\\
			\hline
		\end{tabular}
	\end{table}
	
%	\begin{table}[t]
%		\centering
%		\caption{AC[VI]: H.264/AVC source settings for the non-saturation scenario. Using GOP composed of 3 B frames per I/P frames (G16-B3)~\cite{menth2009source}}
%		\label{tab:video}
%		\begin{tabular}{|c|c|}
%			\hline
%			{\bfseries Parameter} & {\bfseries Value}\\
%			\hline
%			PSNR & $43.5$~dB, best\\
%			GOP size & 16\\
%			GOP & \emph{IBBBPBBBPBBBPBBB}\\
%			Average I size & $5658$~B\\
%			Average P size & $1634$~B\\
%			Average B size & $348$~B\\
%			Frame size standard deviation & 2 times the average\\
%			Average Rate & $300$~kbps\\
%			\hline
%		\end{tabular}
%	\end{table}
	
Apart from the assumptions presented above, the following provide details about the traffic source generators, channel conditions and overall scenarios to be evaluated. Then, simulation results for achieved throughput, number of collisions and time between successful transmissions are presented.

\subsection{Simulation parameters}\label{subsect:simParams}

\subsubsection{Traffic conditions}\label{traffic}
There are two main scenarios regarding traffic generation in a node. The \emph{saturated} traffic condition refers to a node that always has a packet for transmission in its MAC queue. On the other hand, a \emph{non-saturated} node empties its MAC queue and withdraw from the channel contention. These states do not fall far from reality, for instance, a node might be in saturation while it is performing a file transfer. But if instead the node is only performing a VoIP call, its MAC queue will be empty while silence is detected by the codec.

Non-saturation scenarios play an important part on the performance evaluation, specially because both EDCA and CSMA/ECA$_\text{QoS}$ reset their respective CW$_{\text{curr}[AC]}\leftarrow\text{CW}_{\min}[AC]$ when the queue for an specific AC is detected empty, which continuously resets collision-free schedules. Details of the traffic sources for the non-saturated scenario are provided below as well as in Table~\ref{tab:voice}.
	\begin{itemize}
		\item AC[VO] source: we emulate a voice codec with silence detection. That is, when the energy of a voice signal is below a threshold during a determined number of sampled packets, the source stops injecting voice packets into the MAC queue. The Internet Low Bit Rate Codec (iLBC)~\cite{andersen2004internet} is a robust codec designed for IP networks. It features smooth speech quality degradation in case of frame losses, making it suitable for VoIP. It is modelled as an On/Off source, other parameters are shown in Table~\ref{tab:voice}~\cite{menth2009source}. A Constant Bit Rate (CBR) traffic source is active during the On phase.
		
		\item AC[VI] source: follows the characteristics of the H.264/Advanced Video Coding (or H.264/AVC)~\cite{van2008traffic}. Its improved compression tools makes it ideal for high quality video streaming. Video source modelling greatly depends on the video source, that is, action films after packetised produce very different frames than a static interview. This results in rate variability. As also tested in~\cite{van2008traffic}, an example Group of Images (GOP) representative of an action movie source is selected\footnote{Due to its higher rate variability.}. A GOP is composed of I, P and B frames, used to represent past, present and future in a video stream. For a given image quality (PSNR) and size (in pixels by pixels), Table~\ref{tab:voice} shows the average and standard deviation of the I, P and B frame sizes, alongside other video source characteristics.
		
		\item AC[BE] and AC[BK] sources: queues are saturated in all scenarios.
	\end{itemize}
	
As the goal of the \emph{saturated} scenario evaluation is to compare the efficiency of the contention mechanisms used by EDCA and CSMA/ECA$_{\text{QoS}}$, all ACs use circular MAC queues, which are filled at startup with 1470B frames.

\subsubsection{Channel errors}
The inability to receive an ACK frame is handled as a collision, both in EDCA and CSMA/ECA$_{\text{QoS}}$. This could happen due to channel imperfections preventing the receiver from decoding the transmissions. In order to simulate the effects of channel errors over the MAC protocol, we define the likelihood of a MPDU not being acknowledged, $p_e$. It affects every MPDU independently. That is, for every transmission we draw a number from a random variable $X\sim\mathcal{U}[0,1]$, if the number drawn is lower than $p_e$ the frame will not be acknowledged. In the case of MDPU aggregation (AMPDU), it is considered a failed transmission only if all MPDUs in the AMPDU are independently affected by $p_e$. A value of $p_e=0.1$ has been selected for the simulation of the non-saturated scenario, but a comparison with different values for $p_e$ is also provided. The saturation scenario is tested with a perfect channel.

%////////////////////////////////////////////////////////////////////////////////////
%///////////////////////////////////////%Results%//////////////////////////
%///////////////////////////////////////////////////////////////////////////////////

\subsection{CSMA/ECA performance evaluation}\label{sim:results}
This section presents results with Fair Share and TXOP[AC], referred to as CSMA/ECA$_{\text{QoS+FS}}$ and CSMA/ECA$_{\text{QoS+TXOP}}$, respectively. The latter means that upon wining access to the channel an ACs will transmit without contention for as long as indicated by TXOP[AC] in Table~\ref{tab:EDCAparams}. Any kind of frame aggregation is only performed on high priority ACs, that is, AC[VO] and AC[VI]. Furthermore, to provide differentiation between AC[BE] and AC[BK], Schedule Reset is turned off for AC[BK]. This means that $\text{CW}_{\text{curr}}[\text{BK}]$ is only reduced when reaching the retransmission limit or when the queue for this AC is detected empty. In both cases it is reset to $\text{CW}_{\min}[\text{BK}]$.

We first evaluate the performance of CSMA/ECA$_\text{QoS+FS}$, taking special interest to the throughput, failures, fairness and average delay in both traffic conditions. Table~\ref{tab:mac-params} shows the default CSMA/ECA$_\text{QoS}$ settings regarding Schedule Reset, Smart Backoff, and stickiness. Then, we study EDCA and compare the results against CSMA/ECA$_\text{QoS+FS}$, including a mixed network scenario. Next, we replace Fair Share in CSMA/ECA$_\text{QoS}$ with TXOP rules to provide a just comparison with EDCA. We identify this case as CSMA/ECA$_{\text{QoS+TXOP}}$. Finally, we propose a discussion about the results.

\subsubsection{CSMA/ECA$_\text{QoS+FS}$}

	\begin{figure}[t]
	\centering
		\includegraphics[width=0.83\linewidth,angle = -90]{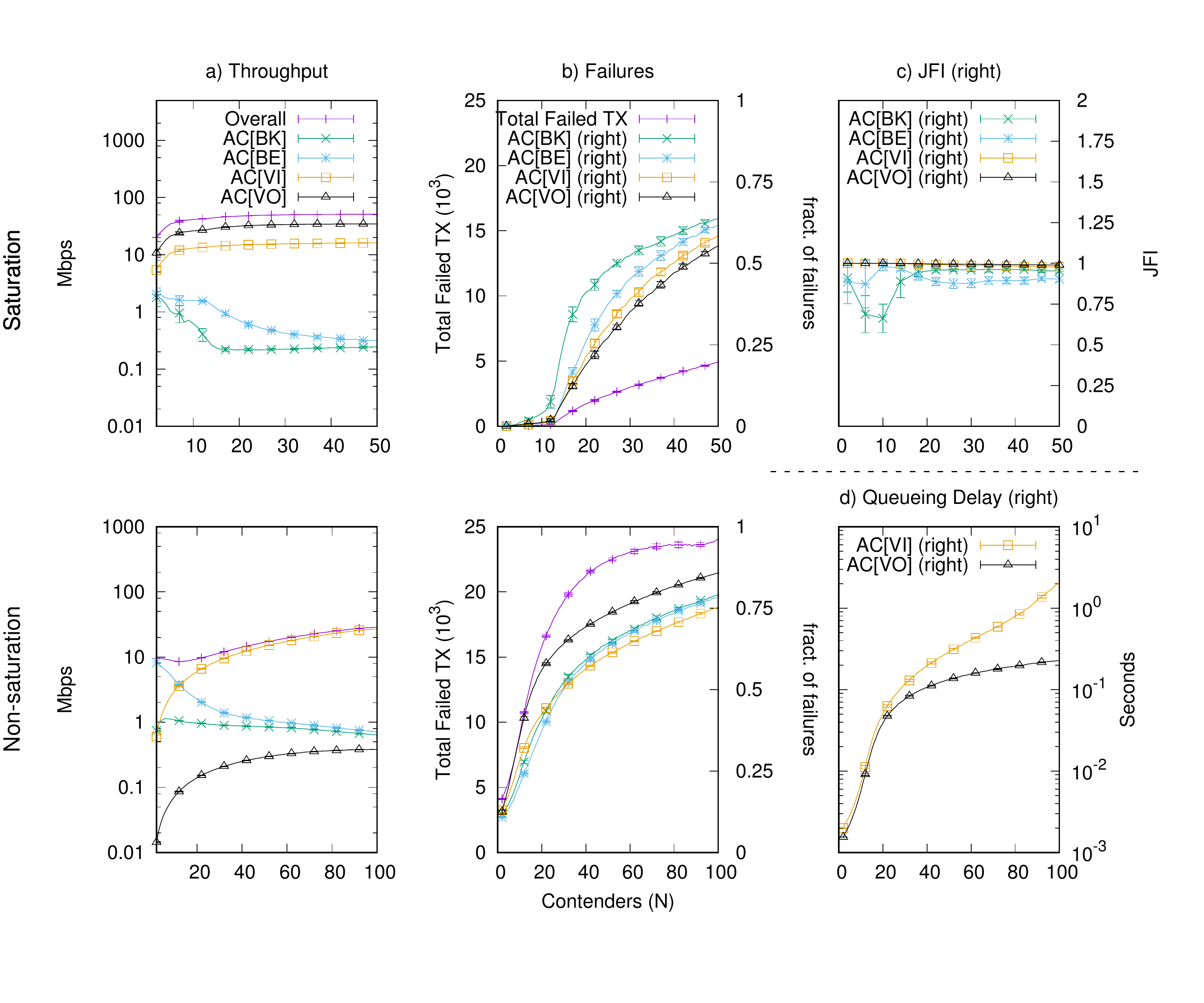}
		\caption{The first row shows: a) Average aggregate throughput, b) failed transmissions, and c) Jain's Fairness Index~\cite{JFI} for CSMA/ECA$_{\text{QoS+FS}}$ in saturation. All frame sizes are equal to 1470B. The bottom row focus on the non-saturation scenario. It shows the same metrics except for the latter, which shows d) the average queueing delay (queue + contention).}
		\label{fig:100}
	\end{figure}

	\begin{figure}[t]
	\centering
		\includegraphics[width=0.4\linewidth,angle = -90]{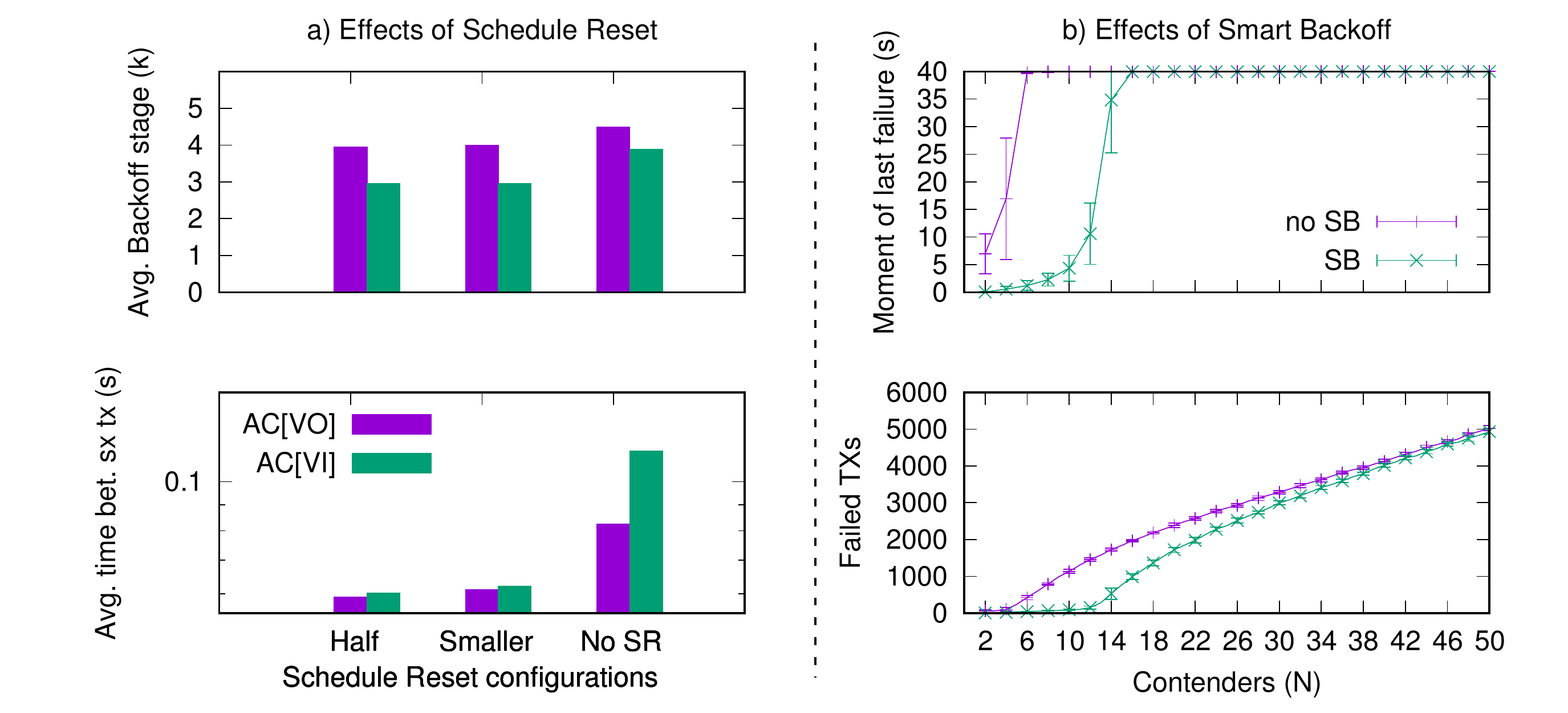}
		\caption{a) Different Schedule Reset configurations with $\gamma=1$ (see Section~\ref{scheduleReset}) in a saturated scenario with $N=8$ contenders. Three SR configurations are proposed. \emph{Half}: SR only attempts to halve the current deterministic backoff; \emph{Smaller}: changes to smaller backoffs are allowed; \emph{no SR}: not using Schedule Reset. b) Shows the effects of the Smart Backoff mechanism. Presenting the moment of the last detected failed transmission, and the total number of failed transmissions.}
		\label{srTest}
	\end{figure}

Figure~\ref{fig:100} shows results from a saturation scenario with a perfect channel on the first row. Columns present a) average throughput, b) failures and c) Jain's Fairness Index (JFI)~\cite{JFI}\footnote{The JFI is an indicator of fairness regarding the ditribution of the available throughput in a system. As the throughput in WLANs is to be equally distributed among contenders, a JFI$=1$ is expected.}. The bottom row of Figure~\ref{fig:100} in turn shows average aggregate throughput, failures, and average queueing delay (queueing + contention time) in non-saturation.

As shown in the figure, CSMA/ECA$_\text{QoS+FS}$ is able to keep a steady overall throughput for a large number of contenders in saturated conditions. Moreover, as ACs aggregate frames proportionally to its current schedule length, throughput fairness is achieved for high priority ACs. Collision-free operation is reached for $N\le12$, as shown in Figure~\ref{fig:100}-b. This is lower than the maximum of $N=32$ mentioned in Section~\ref{ECAqosCollisionFree} and is a consequence of Schedule Reset's $\gamma=1$. For $N\le12$, SR often fails to encounter further reduction opportunities, often succeeding keeping ACs with shorter schedules than the maximum. At higher $N>12$, the aggressiveness of SR due to $\gamma=1$ leads to schedule reductions that cause collisions.

Figure~\ref{srTest} provides a set of comparisons for: a) different configurations of Schedule Reset fixing the number of contenders to $N=8$, and b) the effect of Smart Backoff over CSMA/ECA$_\text{QoS+FS}$ convergence time and failed transmissions in saturated conditions. As shown in Figure~\ref{srTest}-a, the difference between selecting \emph{Half} the current schedule and looking to reduce it to the \emph{Smaller} available length are not significant in terms of average final backoff stage. Nevertheless, a reduction is observed when compared against not using SR. Looking at the average time between successful transmissions, the \emph{Half} configuration provides better results given that a drastic reduction of the schedule increases the collision probability when using $\gamma=1$. As this value of $\gamma$ is required in order to increase the reduction attempts in non-saturated conditions with $p_e>0$, the \emph{Half} configuration is used. That is, Schedule Reset will evaluate the bitmap and only perform a reduction to half the current deterministic backoff.

Smart Backoff prevents virtual collisions and consequent disruption of collision-free schedules. As shown in Figure~\ref{srTest}-b, collision-free operation is only achieved with SB, and for $N\le14$ during simulation time.\footnote{Being the average backoff stage for AC[VO], $k[\text{VO}]=4$, as mentioned in Section~\ref{ECAqosCollisionFree} collision free operation is possible for up to $2^{(k[\text{VO}]-3)}CW_{\min} = 16$ nodes.}.

	\begin{figure}[t]
	\centering
		\includegraphics[width=0.35\linewidth,angle = -90]{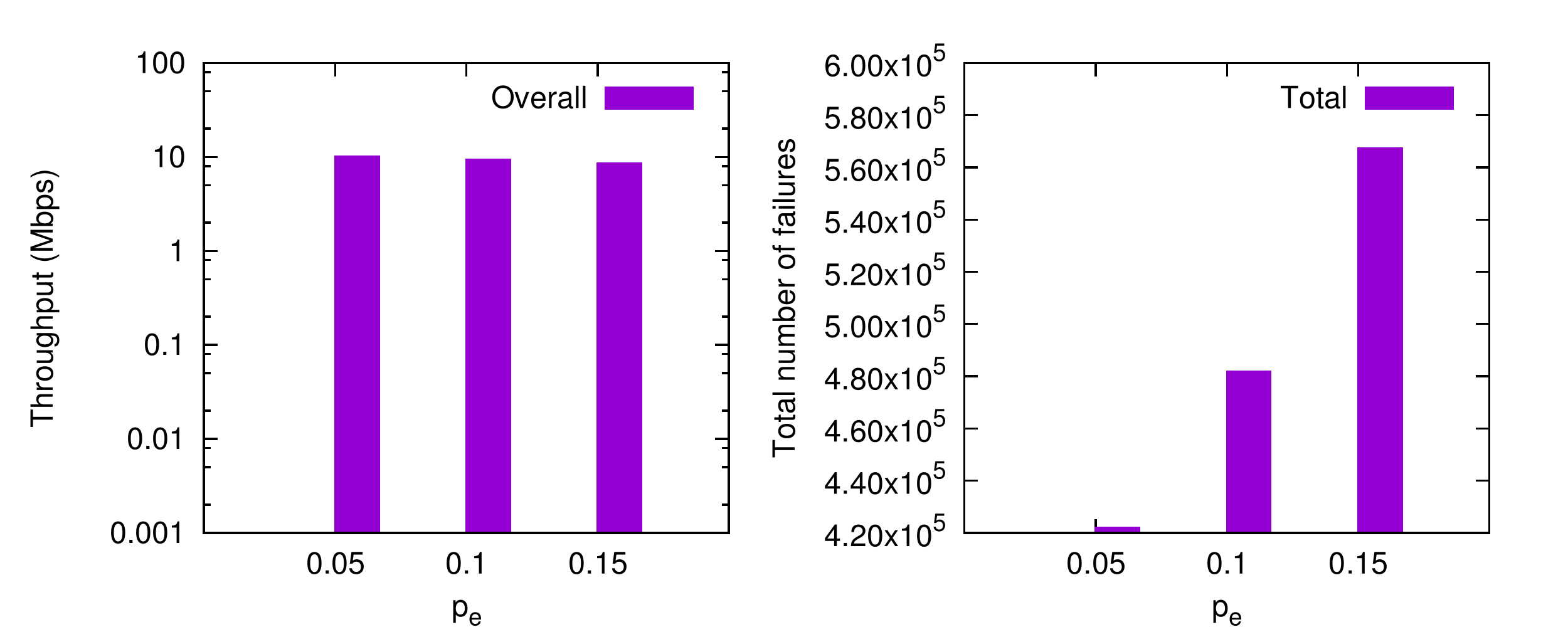}
		\caption{Average throughput and Failed transmissions for different levels of $p_{e}$ in non-saturation with $N=1$.}
		\label{fig:evError}
	\end{figure}

In non-saturation, CSMA/ECA$_{\text{QoS+FS}}$ in Figure~\ref{fig:100} is able to construct collision-free schedules for short periods of time that allow AC[VO] and AC[VI] to saturate at a much higher number of contenders. Further, as shown in Figure~\ref{fig:100}-d the queueing delay of the highest priority AC[VO] is lower than other ACs. The value of $p_e=0.1$ is selected because it produces a moderate increase in the total number of failures observed in Figure~\ref{fig:evError}, where a range of $p_e>0$ with a fixed $N=1$ are tested. As nodes are supposed to be in communication range among each other, we avoid using higher $p_e$ values.

\subsection{EDCA comparison and coexistence}

	\begin{figure}[t]
		\centering
			\includegraphics[width=0.83\linewidth,angle = -90]{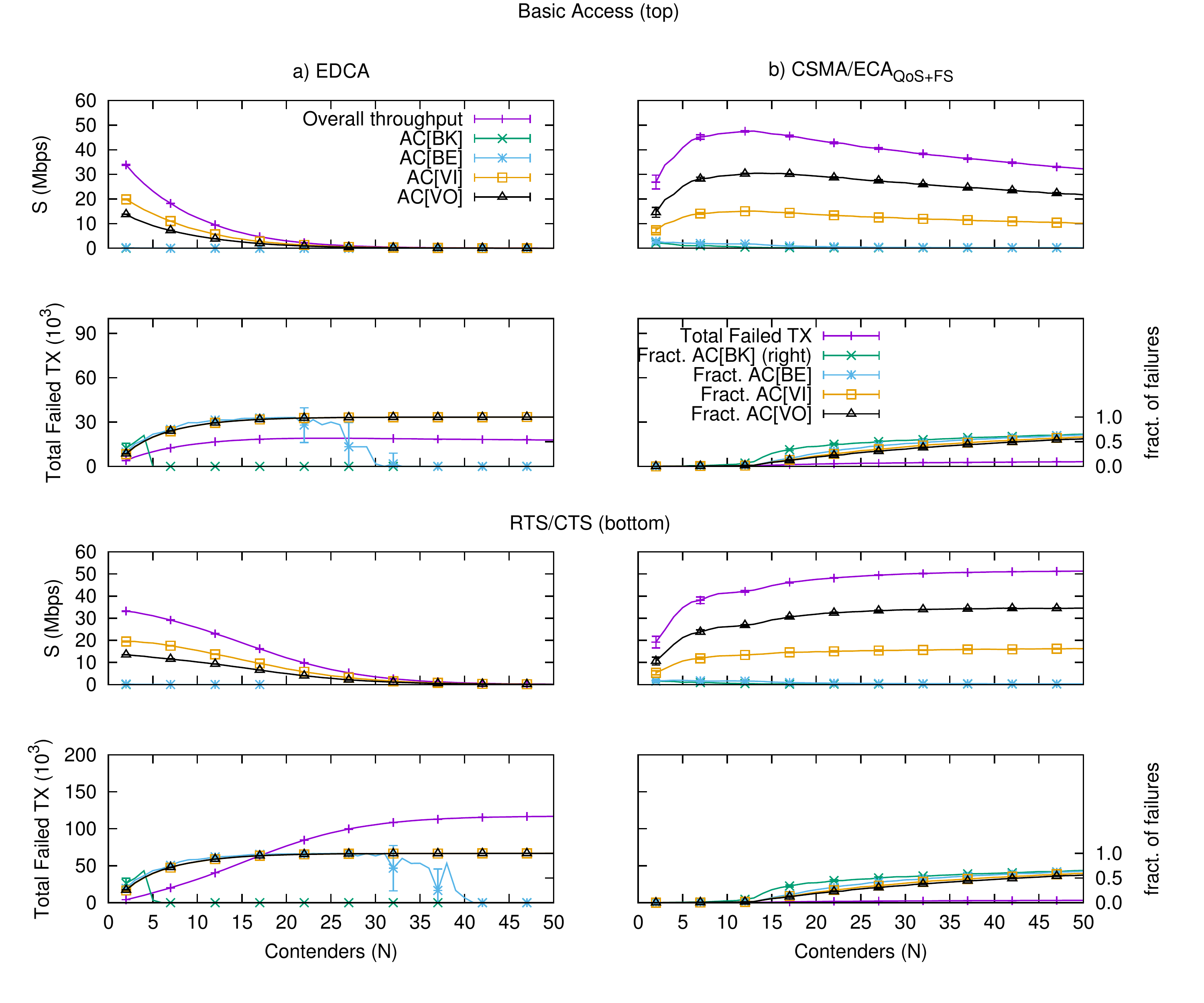}
			\caption{Average aggregate Throughput and Collisions for a) EDCA, and b) CSMA/ECA$_{\text{QoS+FS}}$ in saturation. All frame sizes are equal to 1470B.}
			\label{fig:multiplotSat}
	\end{figure}

Figure~\ref{fig:multiplotSat} gathers the simulation results for average aggregate throughput (S) and failed transmissions in a saturated network using Basic Access (BA) and RTS/CTS. 

EDCA with RTS/CTS (Figure~\ref{fig:multiplotSat}-a bottom) shows higher throughput than using BA. This is an effect of wasting less time recovering from collisions. Moreover, as more time is made available for transmission attempts, RTS/CTS produces a higher total number of failed transmissions, but keeps the same fraction of failures as in BA. RTS/CTS also loosens the starvation of low priority ACs. As indicated by the fraction of failures, the starvation of EDCA AC[BE] occurs at a higher $N=42$, against $N=32$ observed using BA. Given that RTS/CTS is required by the IEEE 802.11e standard when performing frame aggregation, further results do not consider BA.

\begin{figure}[t]
	\centering
		\includegraphics[width=0.35\linewidth,angle = -90]{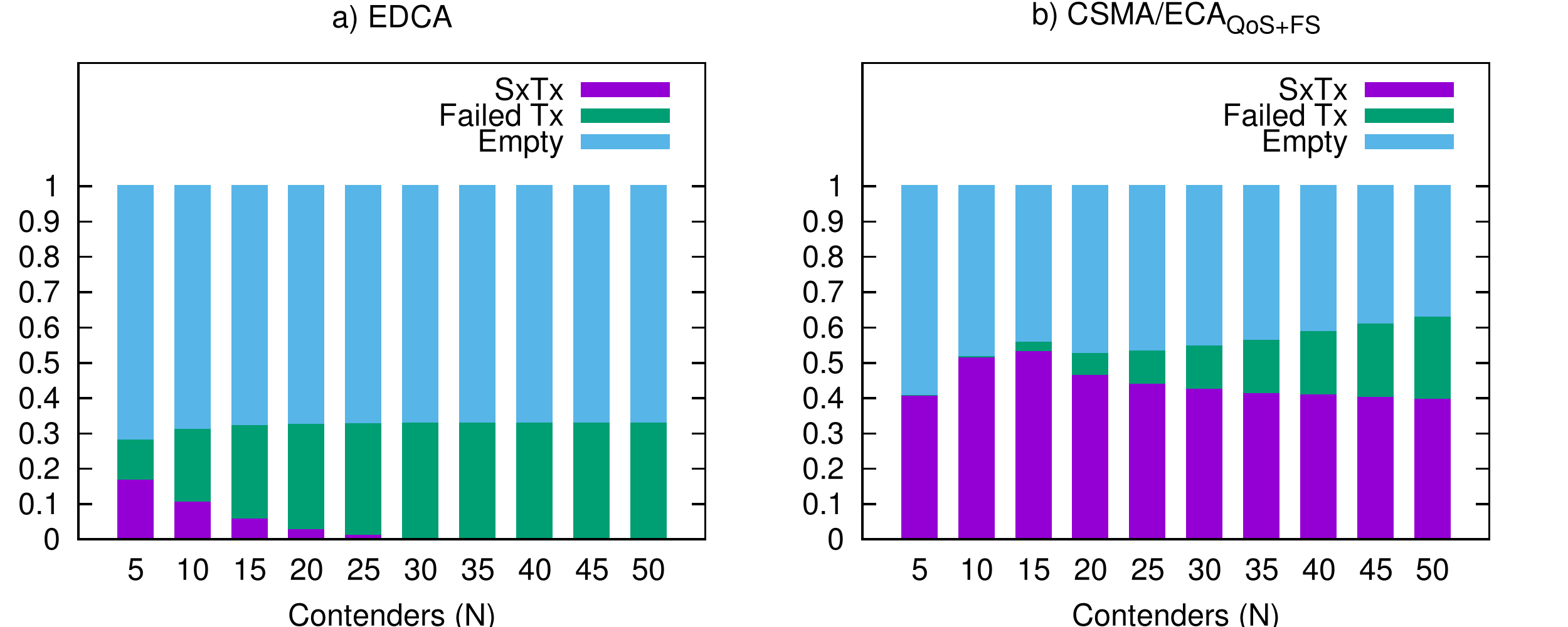}
		\caption{Fraction of slots during a saturated simulation with a growing number of contenders}
		\label{fig:slots}
	\end{figure}

The efficiency of eliminating collisions with CSMA/ECA$_\text{QoS+FS}$ is clearly evident at high number of contenders. Conversely, EDCA's throughput decreases very rapidly, mostly because of an extremely high fraction of failures. Figure~\ref{fig:slots} shows the percentage of empty, successful and failure slots observed during a simulation in saturated conditions.

Despite clearly outperforming EDCA for high number of contenders, CSMA/ECA$_\text{QoS+FS}$ shows lower overall throughput for $N\le5$ in Figure~\ref{fig:multiplotSat}-b. This is due to Fair Share, which aggregates according to the current backoff stage\footnote{That is, $2^{k[AC]}$ frames in an AMPDU (see Section~\ref{scheduleReset}).}. As collisions are quickly eliminated with Smart Backoff, the level of aggregation produced by Fair Share is often lower than TXOP[AC], hence the lower throughput.

	\begin{figure}[t]
	\centering
		\includegraphics[width=0.75\linewidth,angle = -90]{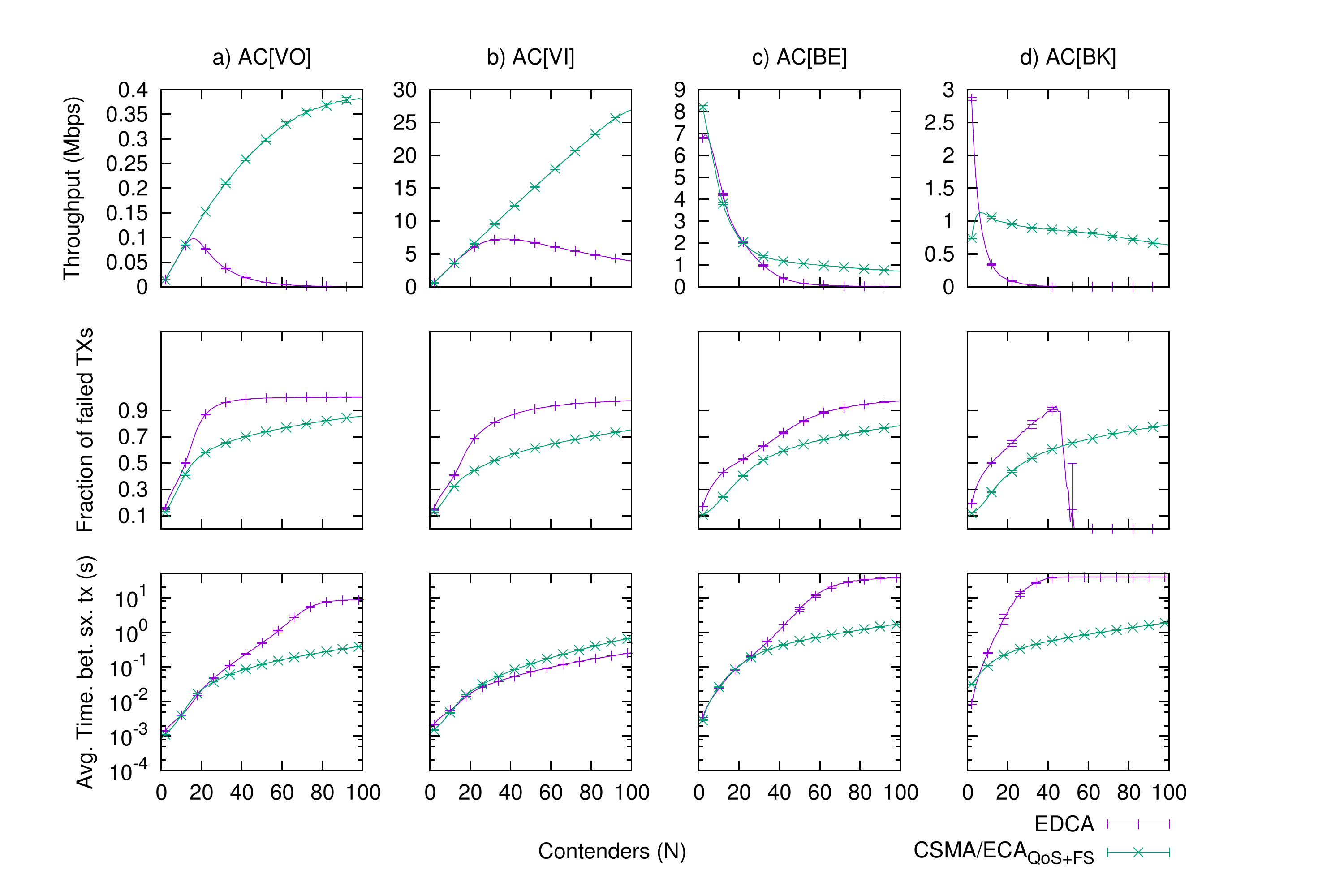}
		\caption{Comparison among protocols per AC in the non-saturation scenario. Each column represents an AC: a) VO, b) VI, c) BE, and d) BK. While rows show: (first) average aggregate throughput, (second) fraction of failed transmissions, and (third) average time between successful transmissions. Legend is located at the bottom right corner of the figure.}
		\label{fig:unsat}
	\end{figure}

Turning to the non-saturation scenario, Figure~\ref{fig:unsat} shows the average aggregate throughput, fraction of failures, and time between successful transmissions as rows $i=(1, 2, 3)$, using labels $j=(a,b,c,d)$ to identify each AC as a column. Subfigures are referred as Figure~\ref{fig:unsat}$.i.j$. 

In Figure~\ref{fig:unsat}.1.a and~\ref{fig:unsat}.1.b, EDCA AC[VO] and AC[VI] achieve less throughput, mainly because they get saturated at lower $N$. CSMA/ECA$_{\text{QoS+FS}}$ AC[VI] on the other hand saturates with a considerable larger $N$\footnote{The average number of aggregated frames using Fair Share is greater than TXOP[VI], thus emptying AC[VI] queue quicker.}. On the other hand, AC[BE] in Figure~\ref{fig:unsat}.1.c shows a slightly higher throughput in EDCA for $6<N\le 16$. This is attributed to the aggressiveness of EDCA's random backoff. Nevertheless, for $N>16$ CSMA/ECA$_{\text{QoS+FS}}$ AC[BE] maintains a steady throughput for an increased number of contenders. Further, CSMA/ECA$_{\text{QoS+FS}}$ AC[BK] outperforms EDCA's for $N>5$ (big deterministic backoffs and the lack of Schedule Reset in AC[BK] account for the lower throughput for $N\le 5$).

%CSMA/ECA$_{\text{QoS+FS}}$ transmissions are longer due to Fair Share, and produce the observed throughput increase. Additionally, these longer transmissions have little effect over the time between successful transmissions of CSMA/ECA$_{\text{QoS+FS}}$ ACs, as shown in Figure~\ref{fig:unsat}.3. This is partly because of the efficient AMPDU aggregation used in 802.11ax~\cite{bellalta2015WCM,IEEE80211ax}.

A big part of CSMA/ECA$_{\text{QoS+FS}}$ throughput enhancement is consequence of a better collision avoidance. This is supported by the reduced fraction of failures shown in Figure~\ref{fig:unsat}.2. Furthermore, the lower fraction of failures observed are the result of the higher saturation point of AC[VO] and AC[VI] due to Fair Share (see Figure~\ref{fig:unsat}.1.a and~\ref{fig:unsat}.1.b).
	
CSMA/ECA$_{\text{QoS+FS}}$ AC[VO] in Figure~\ref{fig:unsat}.3.a has a lower average time between successful transmission, and for a larger number of contenders than EDCA. Looking at AC[VI] in Figure~\ref{fig:unsat}.3.b, CSMA/ECA$_{\text{QoS+FS}}$ shows a higher metric when $N>16$. This is partly due to the duration of successful transmissions of low priority ACs, which are brought to starvation by EDCA as $N$ increases. 

Figure~\ref{fig:unsat}.3.c shows higher time between successful transmissions for EDCA AC[BE]. Conversely, this same metric is slightly higher for CSMA/ECA$_{\text{QoS+FS}}$ AC[BK] at $N\le 5$, as shown in Figure~\ref{fig:unsat}.3.d. As AC[BK] does not use Schedule Reset, the big deterministic backoffs used are responsible for longer periods between successful transmissions. Nevertheless, this effect is reversed for higher $N$.

\subsubsection{Mixed Scenario}\label{mixSection}
\begin{figure}[t]
	\centering
		\includegraphics[width=0.75\linewidth,angle = -90]{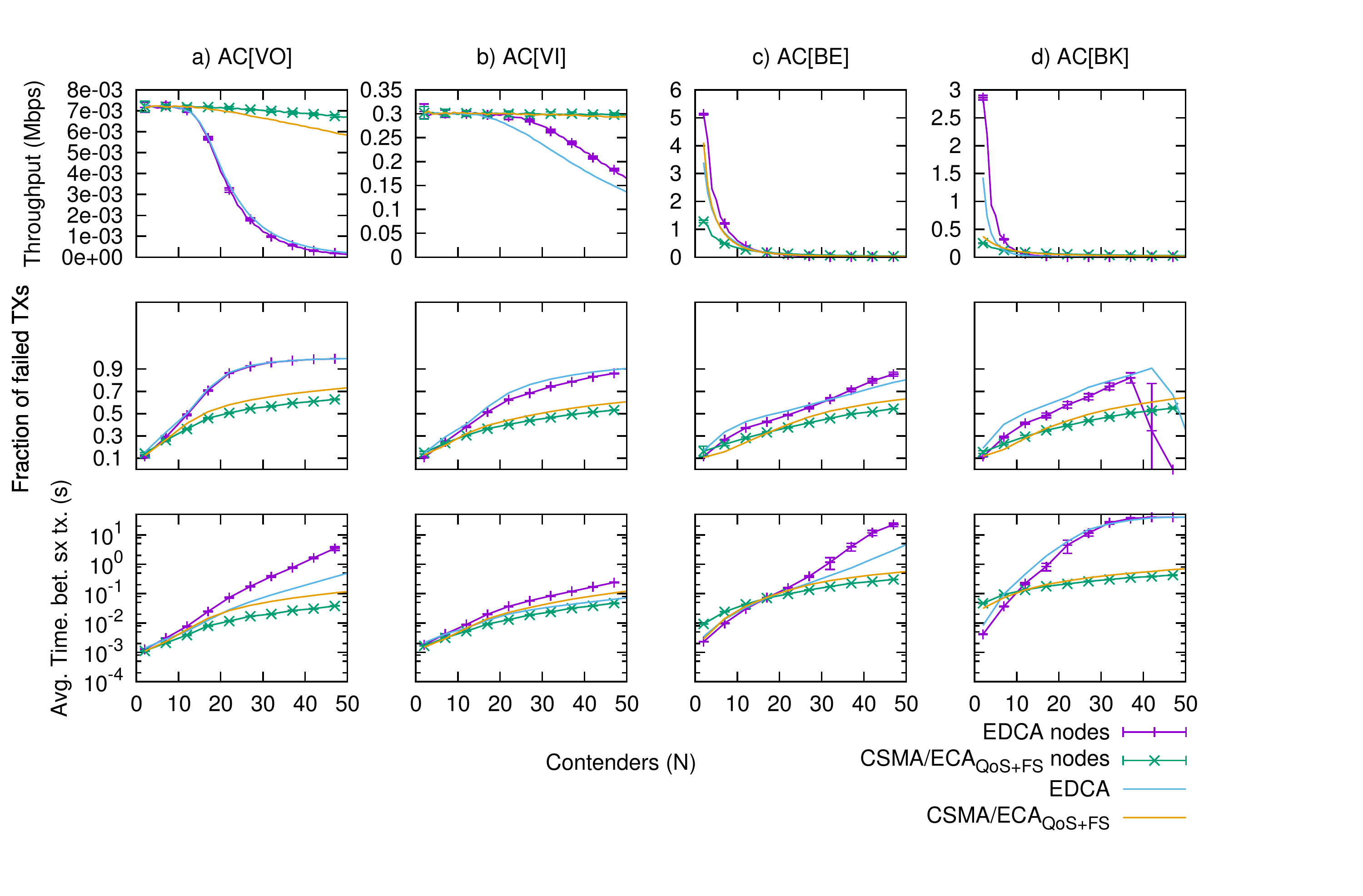}
		\caption{Comparison among protocols per AC in the Mixed Scenario in non-saturation. Each column represents an AC: a) VO, b) VI, c) BE, and d) BK. While rows show: (first) average aggregate throughput per station, (second) fraction of failed transmissions, and (third) average time between successful transmissions. Legend is located at the bottom right corner of the figure.}
		\label{mix}
	\end{figure}

The following results are extracted from simulations performed with a network setup composed of two types of nodes: 50\% EDCA and 50\% CSMA/ECA$_{\text{QoS+FS}}$. It uses the non-saturation scenario settings with $p_e=0.1$. Figure~\ref{mix} shows per node average aggregate throughput, fraction of failed transmissions, and time between successful transmissions as rows $i=(1, 2, 3)$, using labels $j=(a, b, c, d)$ to identify each AC as a column. Subfigures are referred as Figure~\ref{mix}$.i.j$. Curves from pure EDCA and CSMA/ECA$_{\text{QoS+FS}}$ networks are also presented.

Figure~\ref{mix}.1.a shows EDCA AC[VO] getting saturated at around the same number of contenders as in the non-saturated scenario of Figure~\ref{fig:unsat}.1.a ($N=14$). Similarly, as the total number of contenders ($N'$) increases, the throughput of EDCA AC[VO] is degraded even more. CSMA/ECA$_{\text{QoS+FS}}$ nodes are able to avoid collisions more efficiently, resulting in an increased number of successful transmissions for even more users. EDCA AC[VI] in Figure~\ref{mix}.1.b shows the same saturation point as in Figure~\ref{fig:unsat}.1.b.

Still referring to the average throughput, CSMA/ECA$_{\text{QoS+FS}}$ nodes's AC[BE] and AC[BK] in Figure~\ref{mix}.1.c and~\ref{mix}.1.d show lower throughput than EDCA's for $N'_{\text{be}}<16$ and $N'_{\text{bk}}<10$, respectively. This is also observed in Figures~\ref{fig:unsat}.1.c and~\ref{fig:unsat}.1.d. Again, this is because the average deterministic backoff used by CSMA/ECA$_{\text{QoS+FS}}$ AC[BE] and AC[BK] at this number of contenders increases the time between successful transmissions beyond EDCA's. This effect can be seen in Figure~\ref{mix}.3.c and~\ref{mix}.3.d.

Short periods of collision-free operation are achieved among successful CSMA/ECA$_{\text{QoS+FS}}$ ACs due to the use of a deterministic backoff after successful transmissions. This reservation-like\footnote{From the point of view of each AC.}, instead of random contention mechanism is less aggressive, reducing the number of transmission attempts. Nevertheless, it considerably increases efficiency by eliminating collisions. 

Figure~\ref{mix}.3 shows the average time between successful transmissions. EDCA AC[VI] and AC[VO] are negatively affected by CSMA/ECA$_{\text{QoS+FS}}$ nodes. In fact, both ACs's metrics are always higher than CSMA/ECA$_{\text{QoS+FS}}$'s (Figure~\ref{mix}.3.a and~\ref{mix}.3.b). This is mainly due to CSMA/ECA$_{\text{QoS+FS}}$ AC[BE] and AC[BK] transmissions, which are normally starved in crowded EDCA networks.

\subsubsection{CSMA/ECA$_{\text{QoS+TXOP}}$}
Fair Share aggregates up to $32$ frames in an AMPDU, nevertheless, the variable-size video frames proposed for the non-saturation scenario often sum up to more than the maximum TXOP limit defined for EDCA (see Table~\ref{tab:EDCAparams}). Conversely, EDCA aggregates more packets at lower number of nodes. As Fair Share performs aggregation according to the AC's schedule length, at $N\le12$ CSMA/ECA$_\text{QoS+FS}$ ACs reach collision-free operation with short schedules.

\begin{figure}[t]
		\centering
			\includegraphics[width=0.96\linewidth,angle = -90]{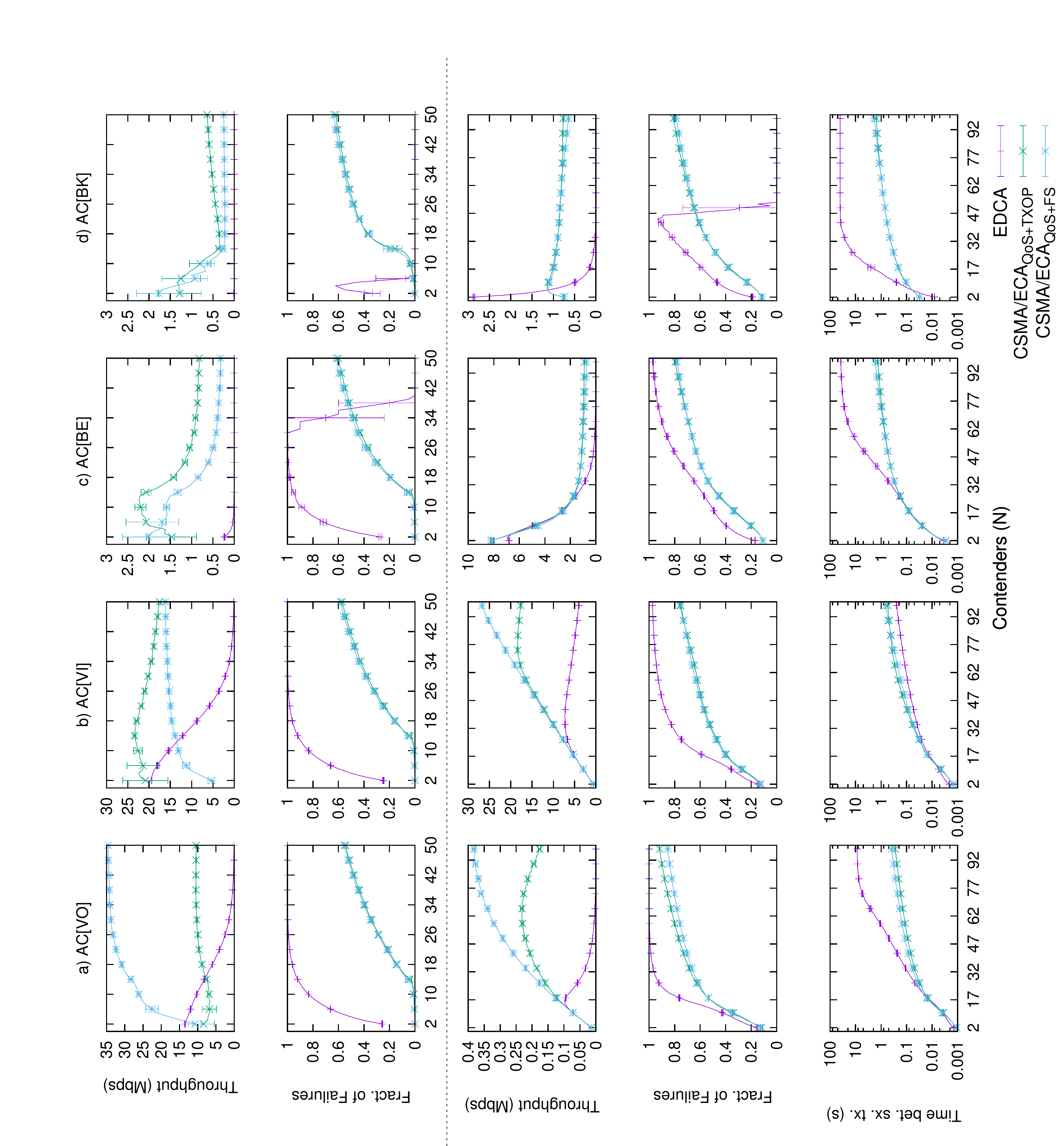}
			\caption{Comparing EDCA, CSMA/ECA$_{\text{QoS+TXOP}}$, and CSMA/ECA$_{\text{QoS+FS}}$ in saturation (top) and non-saturation (bottom). Each column show metrics per AC.}
			\label{fig:multiplotSatTXOP}
	\end{figure}

To provide a just comparison with EDCA, Fair Share is adjusted. That is, AC[VO] and AC[VI] are instructed to always transmit as long as TXOP[AC], as in EDCA. Figure~\ref{fig:multiplotSatTXOP} shows the average aggregate throughput (S) and failed transmissions for EDCA and the adjusted CSMA/ECA$_\text{QoS+TXOP}$ in saturation (top). The bottom of the figure shows the same metrics and the average time between successful transmissions in non-saturation. Columns show the different metrics per AC.

The elimination of collisions with CSMA/ECA$_\text{QoS+TXOP}$ in saturation results in an uneven distribution of the channel resources among contenders for $N\le10$, showing high variability and throughput unfairness. This was originally expected and solved with Fair Share (see Figure~\ref{fig:100}-c and~\cite{sanabria2014high}), but as transmissions are limited by TXOP[AC], ACs with larger schedules are not compensated aggregating more. Instead, CSMA/ECA$_\text{QoS+TXOP}$ ACs pursue opportunities to leverage this issue attempting reductions of the deterministic backoff using Schedule Reset. As the number of contender increases ($N>10$), collisions push all CSMA/ECA$_\text{QoS+TXOP}$ ACs to their largest deterministic backoff, establishing throughput fairness among same category ACs.

CSMA/ECA$_\text{QoS+TXOP}$ ACs rapidly converge to a collision-free operation with Smart Backoff. Results suggest that most of the time high priority ACs, like AC[VO] converge with larger schedules than other low priority ACs. This constitutes a priority inversion in terms of throughput, causing the high variability observed in the first row of Figure~\ref{fig:multiplotSatTXOP} for $N\le10$. 

Looking at the bottom of Figure~\ref{fig:multiplotSatTXOP}, CSMA/ECA$_{\text{QoS+TXOP}}$ clearly outperforms EDCA in the non-saturation scenario, besides, its average time between successful transmissions is practically equal to the one observed in Figure~\ref{fig:unsat}.3.

	\begin{figure}[t]
	\centering
		\includegraphics[width=0.75\linewidth,angle = -90]{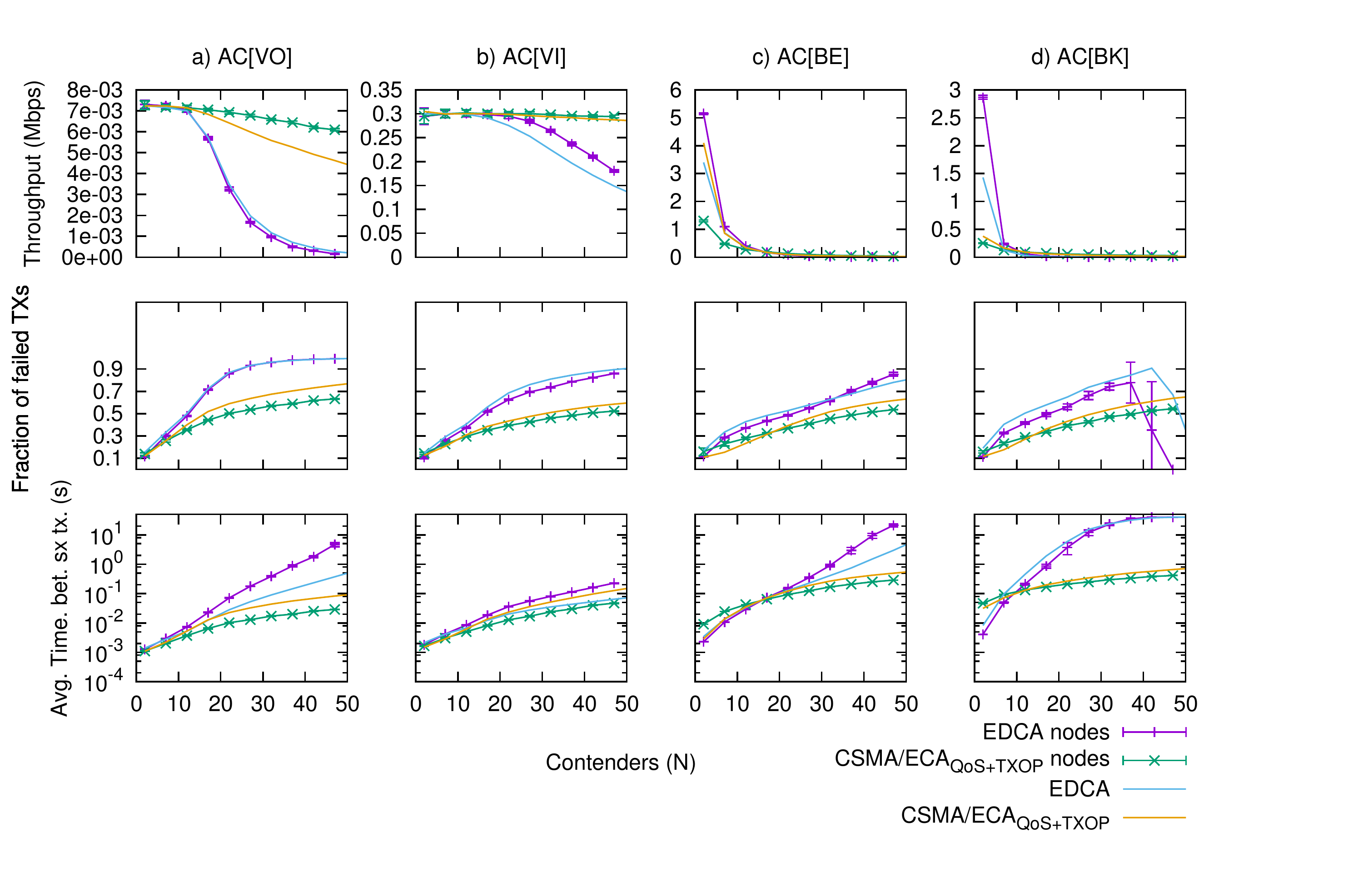}
		\caption{Comparison using CSMA/ECA$_\text{QoS+TXOP}$ in the Mixed Scenario in non-saturation. Each column represents an AC: a) VO, b) VI, c) BE, and d) BK. While rows show: (first) average throughput pero station, (second) fraction of failed transmissions, and (third) average time between successful transmissions. Legend is located at the bottom right corner of the figure.}
		\label{mixTXOP}
	\end{figure}

As Figure~\ref{mix} in Section~\ref{mixSection}, the new Figure~\ref{mixTXOP} shows a Mixed Scenario where 50\% of nodes use EDCA, while the other 50\% use CSMA/ECA$_{\text{QoS+TXOP}}$. The figure shows that the interaction among nodes with different protocols is pretty much the same as when using CSMA/ECA$_{\text{QoS+FS}}$.

	\begin{figure}[t]
	\centering
		\includegraphics[width=0.75\linewidth,angle = -90]{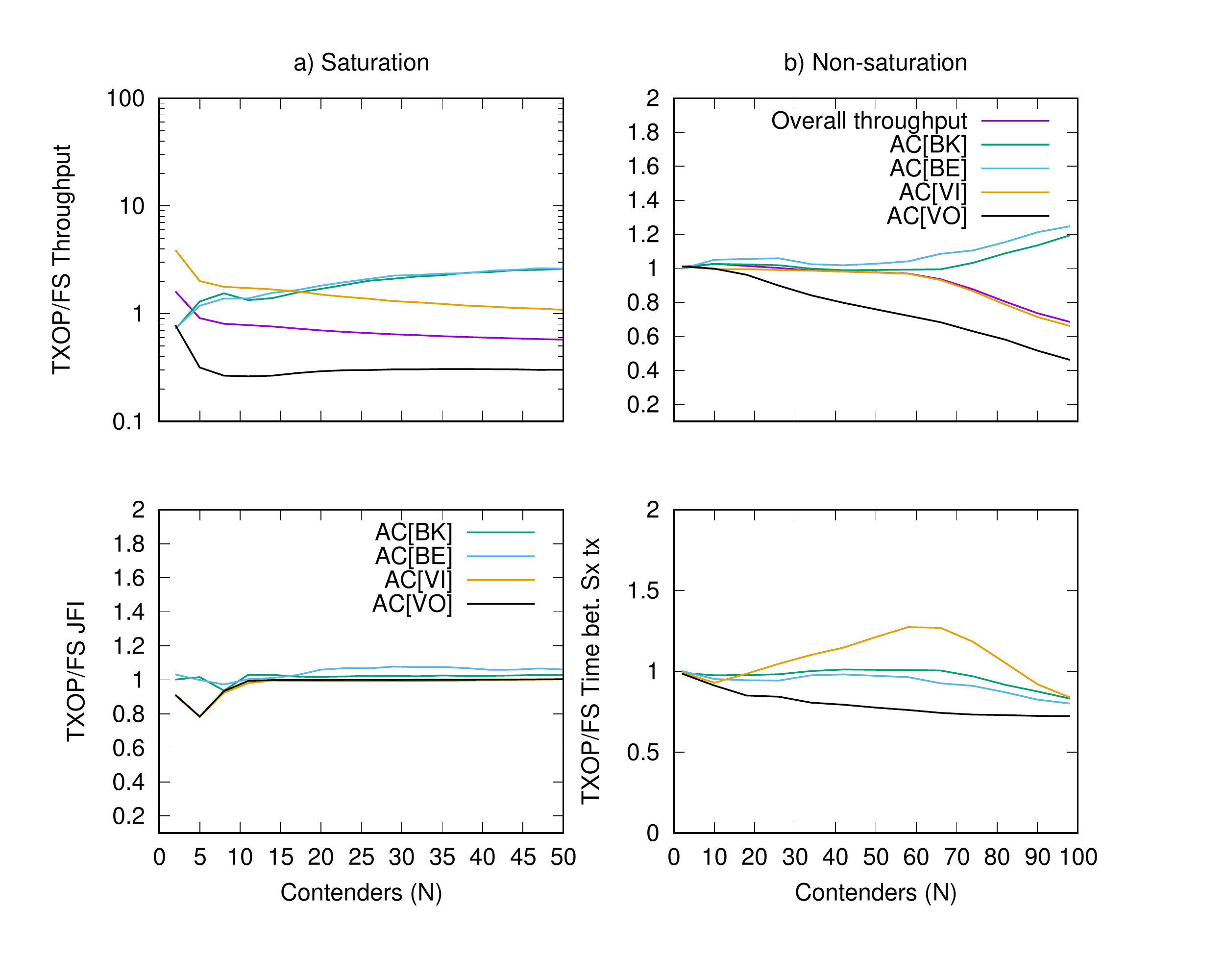}
		\caption{Comparison between TXOP and Fair Share in CSMA/ECA$_\text{QoS}$. Column a) shows the saturation scenario, presenting throughput and JFI. b) presents the non-saturation scenario results, namely throughput and average time between successful transmissions. Results are normalised to CSMA/ECA$_\text{QoS+FS}$ values.}
		\label{txopFS}
	\end{figure}

Figure~\ref{txopFS} shows a comparison between Fair Share and TXOP in CSMA/ECA$_\text{QoS}$. Results are normalised to CSMA/ECA$_\text{QoS+FS}$. The throughput unfairness resulting from using TXOP is clearly appreciable in the saturated scenario for $N\le10$. Nevertheless, CSMA/ECA$_\text{QoS+TXOP}$ shows higher throughput for AC[VI] due to the shorter TXOP[VO]. Figure~\ref{txopFS}-a suggests that the limitation defined by TXOP[VO] allows low priority ACs to achieve higher throughput. This effect is also observed in the non-saturation scenario, referred by Figure~\ref{txopFS}-b. As nodes approach saturation, the shorter TXOP[VO] transmissions  produce an overall reduction in the time between successful transmissions of other ACs.

\subsection{Discussion}
After the analysis, it is clear that the number of contenders, channel and traffic conditions play a main role in the performance of both MAC protocols. 

A perfect channel, RTS/CTS, and low number of contenders are ideal conditions for EDCA in saturation. Nevertheless, CSMA/ECA$_{\text{QoS+TXOP}}$ ACs converge into collision-free schedules with different lengths. Despite Schedule Reset's efforts to reduce the schedule length, ACs rapidly reach collision-free operation and no further reduction is possible without introducing new collisions, producing the throughput oscillations observed in Figure~\ref{fig:multiplotSatTXOP} at $N\le10$. This issue is normally solved with Fair Share. Interestingly, using TXOP aggregation instead of Fair Share produces higher throughput for low number of nodes (despite the irregular throughput distribution), and as TXOP[VO] transmissions are shorter the overall delay of lower priority AC's transmissions is reduced when compared against Fair Share.

As scenarios become crowded with more contenders, CSMA/ECA$_\text{QoS+TXOP}$ consistently outperforms EDCA (see Figure~\ref{fig:multiplotSatTXOP}). Further, it shows lower fraction of failed transmissions for a considerably higher number of contenders. Failed transmissions and non-saturated sources keep changing the structure of Schedule Reset's bitmap, providing more opportunities to reduce the deterministic backoff. Finally, a priority inversion is observed at the bottom row of Figure~\ref{fig:multiplotSatTXOP}, where EDCA AC[VI] shows lower average time between successful transmissions than AC[VO], which is almost starved due to the tight contention parameters.

CSMA/ECA$_\text{QoS+TXOP}$ results suggest it is better than EDCA for crowded scenarios, specially if:
	\begin{itemize}
		\item Traffic differentiation is to be ensured for high number of contenders.
		\item Transmissions from low priority ACs are not to be starved.
		\item To prevent AC priority inversions.
	\end{itemize}

From the point of view of EDCA nodes in the mixed scenario, the deterministic backoff used by the other 50\% of CSMA/ECA$_{\text{QoS+FS}}$ nodes during collision-free periods produce an increase in the number of empty slots. More empty slots imply lower probability of collisions. This means that sharing the network with CSMA/ECA$_{\text{QoS}}$ nodes reduces the collision probability for EDCA nodes. Therefore, the number of successful transmissions from low priority ACs is expected to be higher than in the EDCA-only scenario, increasing the time between successful transmissions of high priority ACs (as shown in Figures~\ref{mix}.3.a,~\ref{mix}.3.b, and Figures~\ref{mixTXOP}.3.a,~\ref{mixTXOP}.3.b).

From the point of view of CSMA/ECA$_{\text{QoS+FS}}$ nodes, the saturation point for AC[VO] and AC[VI] is moved to around $N'=18$, matching EDCA's. Now being saturated, CSMA/ECA$_{\text{QoS+FS}}$ ACs are able to operate without collisions for a number of consecutive transmissions before colliding. This results in a reduction of the time between successful transmissions, coupled with a higher throughput when compared against the non-saturated homogeneous network scenario. The latter still being non-saturated at the same $N=N'$.

\section{Conclusions}\label{section5}
EDCA is able of providing effective traffic differentiation in WLANs. It does so instantiating DCF for each of its four supported MAC queues, or Access Categories (AC), and defining different contention and transmission parameters that allow an statistical differentiation among them. 

Results highlight EDCA's problems at serving many contenders with multiple ACs. Specifically, its contention mechanism being based on a random backoff is in principle unable to eliminate collisions that degrade the overall performance of the network. Strict differentiation techniques like AIFS, and the additional transmission deferrals due to Virtual Collisions starve low-priority ACs in terms of throughput. Further, apart from low priority AC starvation, high priority AC inversion is observed with high number of contenders.

CSMA/ECA$_{\text{QoS}}$ is able to construct collision-free periods that provide an overall throughput increase, while still providing Contention Window-based traffic differentiation for many more contenders. That is, CSMA/ECA$_{\text{QoS}}$ is able to bring traffic differentiation to crowded networks without killing the throughput of low priority ACs, as EDCA does. Further, because both protocols use similar contention parameters, CSMA/ECA$_{\text{QoS}}$ can coexist with EDCA nodes in the same network and still enjoy higher throughput and traffic differentiation.

Authors are even more confident that CSMA/ECA$_\text{QoS}$ can be a suitable replacement for EDCA for high number of contenders because:
	\begin{itemize}
		\item Suppose a simple modification to the existing backoff mechanism of EDCA, and therefore DCF. Suggesting that implementation on real hardware may only require customisation of existing EDCA firmware code, as done with DCF in~\cite{sanabria2014high}.
		\item Is able to support many more high priority flows for a higher number of contenders, making it suitable for the crowded scenarios envisioned for upcoming standard amendments, like 802.11 ax~\cite{IEEE80211ax, bellalta2015WCM}.
		\item Coexistence with EDCA nodes in the same network do not impose costly degradation on the performance. In fact, reduces the collision probability of EDCA nodes allowing them to achieve higher throughput than in an homogeneous network.
	\end{itemize}
	
Even-though our proposal is backwards compatible with EDCA, authors strongly believe in MAC protocol reconfigurability, as done with Wireless MAC Processors using MAClets~\cite{WMP, bianchi2012maclets}. We envision WLANs scenarios where backwards compatibility is no longer an issue because users \emph{download} the MAC protocol from the AP (which can be selected according to different considerations, like: number of users, QoS, privacy, delay, among others). Finally, the most important lesson learned from this work is that there really is no "One-Fits-All" MAC protocol for all WLAN scenarios. We believe that reconfigurability using Software Defined Network-like strategies are the path to follow for future WiFi scenarios.

\section*{ACKNOWLEDGMENT}
This work was partially supported by the Spanish government under project CISNETS (TEC2012-32354) and SGR 2014-1173.

\bibliographystyle{IEEEtran}
\bibliography{ref}

\end{document}